\documentclass[sigconf, 10pt]{acmart}

\usepackage{enumitem}
\usepackage{url}
\usepackage{graphicx}
\usepackage{multirow}
\usepackage{xcolor}
\usepackage{textcomp}
\usepackage{dblfloatfix}    
\usepackage{comment}
\usepackage{hhline}
\usepackage{tabularx,ragged2e}
\usepackage{caption}
\usepackage{subcaption}
\usepackage{algorithm}
\usepackage[noend]{algpseudocode}
\usepackage{lipsum}
\usepackage{wrapfig}
\usepackage{ragged2e}
\usepackage{amsfonts}

\renewcommand\footnotetextcopyrightpermission[1]{} 
\setcopyright{none}

\renewcommand\footnotetextcopyrightpermission[1]{} 
\acmDOI{}

\acmISBN{}

\acmConference[]{Submitted for review to ACM WiNTECH}
\acmYear{2022}

\acmPrice{}

\begin{document}

\title[Designing, Building, and Characterizing\\RF Switch-based Reconfigurable Intelligent Surfaces]{Designing, Building, and Characterizing RF Switch-based Reconfigurable Intelligent Surfaces}

\author{Marco Rossanese}
\affiliation{
	\institution{NEC Laboratories Europe}
}

\author{Placido Mursia}
\affiliation{
	\institution{NEC Laboratories Europe}
}

\author{Andres Garcia-Saavedra}
\affiliation{
	\institution{NEC Laboratories Europe}
}

\author{Vincenzo Sciancalepore}
\affiliation{
	\institution{NEC Laboratories Europe}
}

\author{Arash Asadi}
\affiliation{
	\institution{TU Darmstadt, Germany}
}

\author{Xavier Costa-Perez}
\affiliation{
	\institution{i2cat, NEC Labs Europe and ICREA}
}
\renewcommand{\shortauthors}{Rossanese \emph{et al.}}

\begin{abstract}

In this paper, we present our experience designing, prototyping, and empirically characterizing RF Switch-based Reconfigurable Intelligent Surfaces (RIS).
Our RIS design comprises arrays of patch antennas, delay lines and programmable radio-frequency (RF) switches that enable passive 3D beamforming, i.e., without active RF components. 
We implement this design using PCB technology and low-cost electronic components, and thoroughly validate our prototype in a controlled environment with high spatial resolution codebooks. 
%
%
Finally, we make available a large dataset with a complete characterization of our RIS and present the costs associated with reproducing our design.

\end{abstract}

\maketitle
\newcommand{\bs}[1]{\boldsymbol{#1}}%
\newcommand{\roundP}[1]{\left(#1\right)}%
\newcommand{\squareP}[1]{\left[#1\right]}%
\newcommand{\normS}[1]{\norm{#1} ^2}%
\newcommand{\smo}[1]{\smashoperator[r]{#1}}%
\newcommand{\overbar}[1]{\mkern 1.5mu\overline{\mkern-1.5mu#1\mkern-1.5mu}\mkern 1.5mu}
\newcommand{\rmm}[1]{\mathrm{#1}}
\newcommand{\red}[1]{\textcolor{red}{#1}}
\newcommand{\st}{\mathrm{subject~to}}

\makeatletter
\let\oldabs\abs
\def\abs{\@ifstar{\oldabs}{\oldabs*}}

\let\oldnorm\norm
\def\norm{\@ifstar{\oldnorm}{\oldnorm*}}
\makeatother
\renewcommand{\a}{\mathbf{a}}
\renewcommand{\b}{\mathbf{b}}
\renewcommand{\c}{\mathbf{c}}
\renewcommand{\d}{\mathbf{d}}
\newcommand{\e}{\mathbf{e}}
\newcommand{\f}{\mathbf{f}}
\newcommand{\g}{\mathbf{g}}
\newcommand{\h}{\mathbf{h}}
\renewcommand{\i}{\mathbf{i}}
\renewcommand{\j}{\mathbf{j}}
\renewcommand{\k}{\mathbf{k}}
\newcommand{\m}{\mathbf{m}}
\newcommand{\n}{\mathbf{n}}
\renewcommand{\o}{\mathbf{o}}
\newcommand{\p}{\mathbf{p}}
\newcommand{\q}{\mathbf{q}}
\renewcommand{\r}{\mathbf{r}}
\newcommand{\s}{\mathbf{s}}
\renewcommand{\t}{\mathbf{t}}
\renewcommand{\u}{\mathbf{u}}
\renewcommand{\v}{\mathbf{v}}
\newcommand{\w}{\mathbf{w}}
\newcommand{\x}{\mathbf{x}}
\newcommand{\y}{\mathbf{y}}
\newcommand{\z}{\mathbf{z}}

\newcommand{\0}{\mathbf{0}}
\newcommand{\1}{\mathbf{1}}

\newcommand{\A}{\mathbf{A}}
\newcommand{\B}{\mathbf{B}}
\newcommand{\D}{\mathbf{D}}
\newcommand{\E}{\mathbf{E}}
\newcommand{\F}{\mathbf{F}}
\renewcommand{\H}{\mathbf{H}}
\newcommand{\I}{\mathbf{I}}
\newcommand{\J}{\mathbf{J}}
\newcommand{\K}{\mathbf{K}}
\renewcommand{\L}{\mathbf{L}}
\newcommand{\M}{\mathbf{M}}
\newcommand{\N}{\mathbf{N}}
\renewcommand{\O}{\mathbf{O}}
\renewcommand{\P}{\mathbf{P}}
\newcommand{\Q}{\mathbf{Q}}
\newcommand{\R}{\mathbf{R}}
\newcommand{\T}{\mathbf{T}}
\newcommand{\V}{\mathbf{V}}
\newcommand{\W}{\mathbf{W}}
\newcommand{\X}{\mathbf{X}}
\newcommand{\Y}{\mathbf{Y}}
\newcommand{\Z}{\mathbf{Z}}

\newcommand{\alphab}{\boldsymbol{\alpha}}
\newcommand{\betab}{\boldsymbol{\beta}}
\newcommand{\gammab}{\boldsymbol{\gamma}}
\newcommand{\deltab}{\boldsymbol{\delta}}
\newcommand{\epsilonb}{\boldsymbol{\epsilon}}
\newcommand{\varepsilonb}{\boldsymbol{\varepsilon}}
\newcommand{\zetab}{\boldsymbol{\zeta}}
\newcommand{\etab}{\boldsymbol{\eta}}
\newcommand{\thetab}{\boldsymbol{\theta}}
\newcommand{\varthetab}{\boldsymbol{\vartheta}}
\newcommand{\iotab}{\boldsymbol{\iota}}
\newcommand{\kappab}{\boldsymbol{\kappa}}
\newcommand{\lambdab}{\boldsymbol{\lambda}}
\newcommand{\mub}{\boldsymbol{\mu}}
\newcommand{\nub}{\boldsymbol{\nu}}
\newcommand{\xib}{\boldsymbol{\xi}}
\newcommand{\pib}{\boldsymbol{\pi}}
\newcommand{\varpib}{\boldsymbol{\varpi}}
\newcommand{\rhob}{\boldsymbol{\rho}}
\newcommand{\varrhob}{\boldsymbol{\varrho}}
\newcommand{\sigmab}{\boldsymbol{\sigma}}
\newcommand{\varsigmab}{\boldsymbol{\varsigma}}
\newcommand{\taub}{\boldsymbol{\tau}}
\newcommand{\upsilonb}{\boldsymbol{\upsilon}}
\newcommand{\phib}{{\boldsymbol{\phi}}}
\newcommand{\varphib}{{\boldsymbol{\varphi}}}
\newcommand{\chib}{\boldsymbol{\chi}}
\newcommand{\psib}{\boldsymbol{\psi}}
\newcommand{\omegab}{\boldsymbol{\omega}}

\newcommand{\Gammab}{\mathbf{\Gamma}}
\newcommand{\Deltab}{\mathbf{\Delta}}
\newcommand{\Thetab}{\mathbf{\Theta}}
\newcommand{\Lambdab}{\mathbf{\Lambda}}
\newcommand{\Xib}{\mathbf{\Xi}}
\newcommand{\Pib}{\mathbf{\Pi}}
\newcommand{\Sigmab}{\mathbf{\Sigma}}
\newcommand{\Upsilonb}{\boldsymbol{\Upsilon}}
\newcommand{\Phib}{\mathbf{\Phi}}
\newcommand{\Psib}{\mathbf{\Psi}}
\newcommand{\Omegab}{\mathbf{\Omega}}

\newcommand{\setA}{\mathcal{A}}
\newcommand{\setB}{\mathcal{B}}
\newcommand{\setC}{\mathcal{C}}
\newcommand{\setD}{\mathcal{D}}
\newcommand{\setE}{\mathcal{E}}
\newcommand{\setF}{\mathcal{F}}
\newcommand{\setG}{\mathcal{G}}
\newcommand{\setH}{\mathcal{H}}
\newcommand{\setI}{\mathcal{I}}
\newcommand{\setJ}{\mathcal{J}}
\newcommand{\setK}{\mathcal{K}}
\newcommand{\setL}{\mathcal{L}}
\newcommand{\setM}{\mathcal{M}}
\newcommand{\setN}{\mathcal{N}}
\newcommand{\setO}{\mathcal{O}}
\newcommand{\setP}{\mathcal{P}}
\newcommand{\setQ}{\mathcal{Q}}
\newcommand{\setR}{\mathcal{R}}
\newcommand{\setS}{\mathcal{S}}
\newcommand{\setT}{\mathcal{T}}
\newcommand{\setU}{\mathcal{U}}
\newcommand{\setV}{\mathcal{V}}
\newcommand{\setW}{\mathcal{W}}
\newcommand{\setX}{\mathcal{X}}
\newcommand{\setY}{\mathcal{Y}}
\newcommand{\setZ}{\mathcal{Z}}

\newcommand{\Real}{\mbox{$\mathbb{R}$}}
\newcommand{\Compl}{\mbox{$\mathbb{C}$}}
\newcommand{\VI}{\mathrm{VI}}

\newcommand{\argmin}{\operatornamewithlimits{argmin}}
\newcommand{\argmax}{\operatornamewithlimits{argmax}}
\newcommand{\diag}{\mathrm{diag}}
\newcommand{\Diag}{\mathrm{Diag}}
\newcommand{\diff}{\mathrm{d}}
\newcommand{\Exp}{\mathbb{E}}
\newcommand{\rmF}{\mathrm{F}}
\newcommand{\herm}{\mathrm{H}}
\renewcommand{\Im}{\mathrm{Im}}
\renewcommand{\Pr}{\mathbb{P}}
\newcommand{\rank}{\mathrm{rank}}
\renewcommand{\Re}{\mathrm{Re}}
\newcommand{\tr}{\mathrm{tr}}
\newcommand{\tran}{\mathrm{T}}
\newcommand{\Var}{\mathrm{Var}}

\section{Introduction} 

Reconfigurable Intelligent Surfaces (RISs)
are well-perceived as a key technology for next-generation mobile systems~\cite{9475160}.
Despite the recent hype on the topic, mostly driven by theoretical models and simulations, empirical studies are scarce due to lack of accessible and affordable RIS prototypes.

A RIS is essentially a planar structure with passive reflective cells that can control the electromagnetic response of impinging radio-frequency (RF) signals, such as changes in phase, amplitude, or polarization. 
%
%
Indeed, RISs open a new paradigm~\cite{albanese22} where the wireless channel---traditionally treated simply as an optimization constraint---plays an active role subject to optimization with the potential of increasing the energy efficiency of mobile networks by $>\!\!50\%$~\cite{9497709}. 

To this end, a RIS must satisfy the following requirements:
    ($i$) RISs shall (re-)steer RF signals with minimal power loss;
    ($ii$) RISs must not use active RF components; 
    ($iii$) RISs must minimize the energy required to re-configure their reflective cells;
    ($iv$) RISs shall be re-configurable in real-time; and
    ($v$) RISs must be amenable to low-cost production at scale.

{\bf Related Work.}  \label{sec:related}
Among existing literature on prior experiences, 
%
\cite{yezhen2020novel} discloses a $16$x$16$ RIS operating at $28$GHz, whereas \cite{trichopoulos2021design} introduces a $16$x$10$ RIS working at sub-$6$GHz with an Arduino control unit where only groups of elements can be configured. Both solutions implement a PIN diode-based RIS with a $1$-bit resolution phase shift. Additionally, the prototype presented in \cite{dai2020reconfigurable} obtains a $2$-bit phase quantization by using $5$ PIN diodes per RIS element, whereas in \cite{VincentPoor2020} the use of $3$ PIN diodes allows for $8$ phase states. 
\cite{RomainDiRenzo2021} discloses a $14$x$14$ RIS based on varactor diodes, which allows continuous control of the phase shifts at the cost of a wide range in control voltages, which is generally hard to achieve. 

Conversely, RF switch-based implementations unveil a lower cost w.r.t. PIN diodes used to control the reflector units. In particular  
in~\cite{tan2018enabling}, a RIS prototype with $14$x$16$ reflectors at $60$GHz is presented. The unit elements are placed more than one $\lambda$ away to reduce the mutual coupling at the expense of a tighter maximum scanning angle ($\lambda$ is the operating wavelength). 
%
In~\cite{arun2020rfocus}, $40$ reflectors are mounted on the boards, they are $\lambda/4$ tall, $\lambda/10$ wide, and separated $\lambda/10$ on both the x and y-axis. 
Finally, \cite{scattermimo} allows singular configurations of the elements, thereby reducing the number of utilized pins at controller side. This RIS is made of $4$x$4$ patch antennas operating at $5$GHz and controlled with $2$-bit phase shifter made with the transmission line method. 

{\bf Our design.} Besides meeting the usual requirements for a RIS, our approach provides additional features compared to previous work. First, in contrast to diode-based approaches, usually constrained to 1-bit phase shifters, our design has a resolution of 3 bits, enabling \emph{high spatial resolution codebooks}. 

Second, our design permits coordinating multiple ($M$) boards, and hence it enables \emph{larger-scale structures} of $M \times N_x \times N_y$ surfaces in a modular and flexible manner. 

Finally, in addition to producing phase shifts onto impinging signals in a programmable manner, we can configure individual cells to fully absorb the energy of RF signals, which open several opportunities. For instance, we can effectively switch off reflective components, which let us \emph{virtually optimize the shape and the size of the RIS} to meet system constraints. 
We would like to remark that \emph{switching off} a cell in a RIS is not trivial when a cell is specifically designed to reflect signals passively, without a direct energy feed that could be cut off.
%



\section{RIS design}\label{sec:design}

\begin{figure}
\vspace{-2mm}
    \centering
    \includegraphics[width=0.7\linewidth]{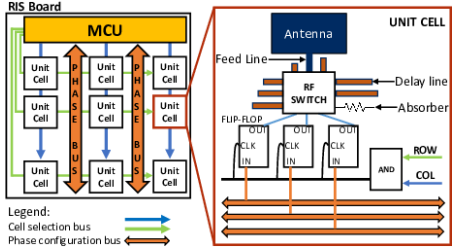}
    \vspace{-3mm}
    \caption{\small RIS board and unit cell.}
    \label{fig:unit_cell}
    \label{fig:ris_board}
\end{figure}

The main purpose of our RIS is to perform 3D beamforming passively, i.e., re-focus the energy received from impinging RF signals towards specified directions without active (energy-consuming) RF components. 
Fig.~\ref{fig:ris_board} illustrates a \emph{\bf board} consisting of a grid of $N_x \times N_y$ \emph{\bf unit cells} distributed in a 2D array. Unit cells are elements that can reflect RF signals with configurable phase shifts. 
%
%

Phase shifts are configured by a microcontroller unit (MCU), which can be programmed from an external controller.
%
%
%
Conventional RIS designs are characterized by dedicated $N_x \times N_y$ connections from the MCU to each unit cell. However, MCUs only support a limited number of such connections, which constrains the maximum number of cells and, consequently, the achievable beamforming gains~\cite{RISMA}.
A more scalable approach is to connect each cell with a pair of buses, denoted as \emph{\bf column/row cell selection buses}, that \emph{select} the cell to be configured, and a \emph{\bf phase configuration bus}, which communicates the desired configuration index out of a discrete set. In this way,  we reduce the complexity of the design from $N_x \times N_y$ to $N_x + N_y$ connections per board.

As shown in the right-hand side of Fig.~\ref{fig:ris_board}, each unit cell connects both column/row selection buses with an \emph{\bf AND} gate. Hence, when the MCU sets a high voltage state in row $x$ and column $y$, the MCU activates the configuration bus for unit cell $(x,y)$, whereas all the remaining gates across the board will output a low voltage state ($0$~V).
%
Each cell also integrates a set of \emph{\bf flip-flop D}, which exploit the high-state exiting the AND gate as a rising edge to update and send out the value stored in memory.
We designed our RIS with 3-bit phase shifters, which enable high spatial resolution codebooks. Therefore, each cell uses \emph{three} 1-bit phase configuration buses and \emph{three} flip-flops. 

The latter are connected to \emph{\bf configuration ports} in an \emph{\bf RF switch}. An RF switch is a component that can redirect the RF signal received from an \emph{\bf input port} towards one \emph{\bf output port}, as indicated by the configuration ports. The input port is connected to a \emph{\bf patch antenna}, the ultimate responsible for interacting with the medium, through a \emph{\bf feeding line}.
%
Each output port (except one) uses an open-ended \emph{\bf delay line} with a suitably-designed length to reflect impinging signals with a specific time delay, shifting the signal phase. 

\begin{figure}[t!]
\vspace{-2mm}
\minipage{0.5\columnwidth}
\centering
  \includegraphics[width=\columnwidth]{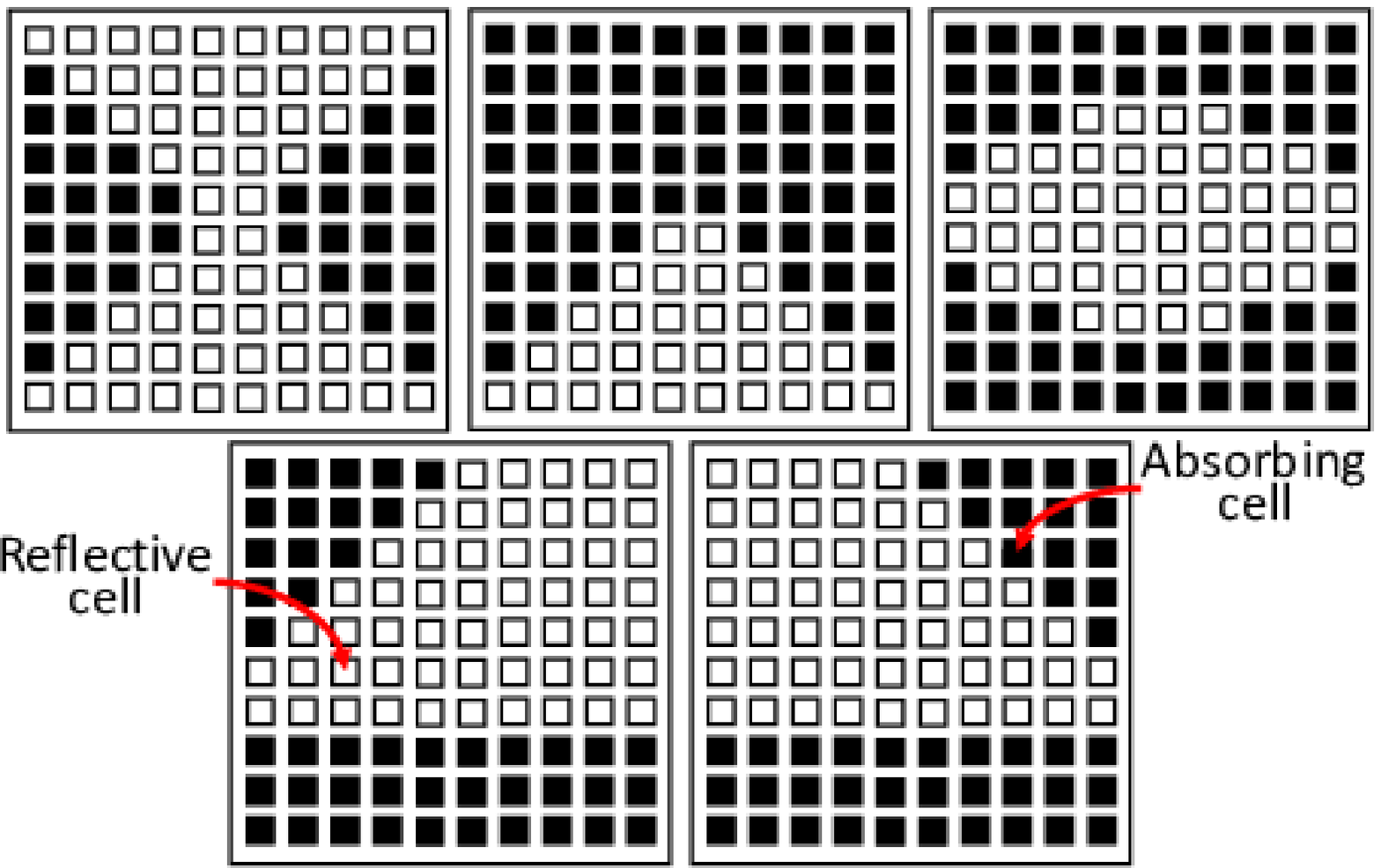}
    \vspace{-6mm}
  \caption{\small Shape-adaptive RIS.}
  \label{fig:adaptive_shapes}
\endminipage
\hfill
\minipage{0.49\columnwidth}
\centering
  \includegraphics[width=\columnwidth]{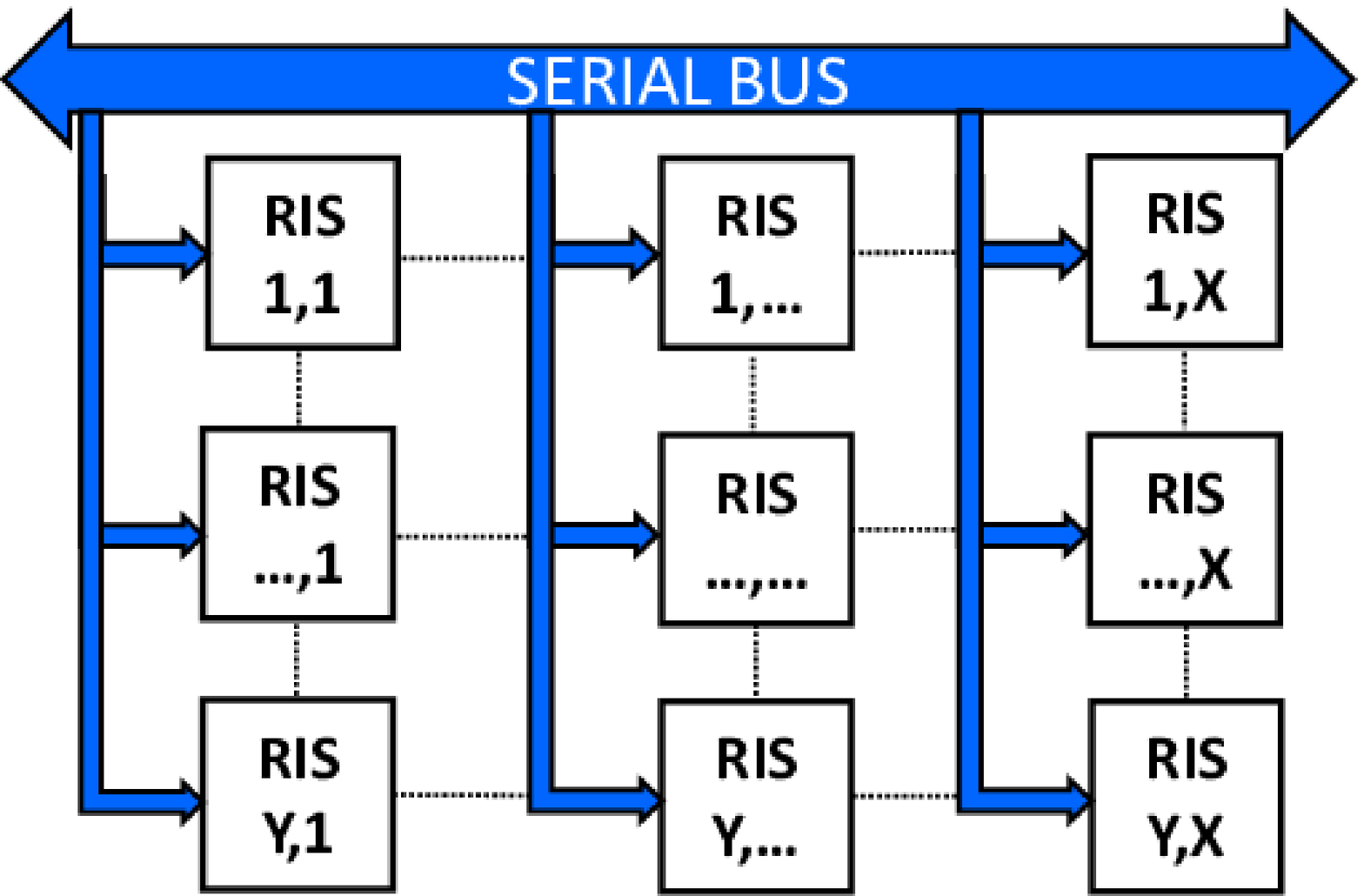}
  \vspace{-7mm}
  \caption{\small Multi-board RIS.}
  \label{fig:ris_all}
\endminipage
\end{figure}

We reserve one output port of the RF switch to connect an \emph{\bf absorber}, an impedance-matching component that absorbs the energy of incoming signals instead of reflecting them back. We call this configuration \emph{absorption state}, and it let us virtually optimize the reflective area of the RIS to meet system constraints. For instance, we can flexibly adapt to different time constraints when optimizing the RIS configuration (which takes longer the larger the number of active cells in the RIS. This is illustrated in Fig.~\ref{fig:adaptive_shapes}.
%
%
Alternatively, an energy harvester~\cite{mir2021passivelifi} may be employed instead to re-use the dissipated energy to feed a low-consuming MCU, becoming self-sustainable boards, which we leave for future work.

Our RIS design is modular: as shown in Fig.~\ref{fig:ris_all}, multiple boards can be coordinated through a common bus. The disposition of the unit cells across different boards has been carefully designed to have a separation of $\lambda/2$, where $\lambda$ is the operating wavelength. Such modular boards let us increase/decrease the physical area of our structure without compromising the inter-cell distance, as depicted in Fig.~\ref{fig:adaptive_shapes}.

\vspace{-1mm}
\section{Beamforming codebook}\label{sec:model}

A RIS board can be modelled as a uniform planar array (UPA) comprised of $N=N_x N_y$ antenna elements~\cite{basar2019wireless, RISMA}. Hence, we define the array response at the RIS for the steering angles $\bar{\theta}$, $\bar{\phi}$ along the azimuth and elevation, respectively, as
\begin{align}
    \a(\bar{\theta},\bar{\phi}) & \triangleq \a_x(\bar{\theta}) \otimes \a_y(\bar{\phi}) \label{eq:UPA}  \\
    & = [1,e^{j2\pi\delta \cos(\bar{\theta})},\ldots,e^{j2\pi\delta(N_x-1) \cos(\bar{\theta})}] \nonumber \\
    & \otimes [1,e^{j2\pi\delta \sin(\bar{\phi})},\ldots,e^{j2\pi\delta(N_y-1) \sin(\bar{\phi})}] \in \Compl^{N\times 1} , \nonumber
\end{align}
where $\delta$ is the ratio between the antenna spacing and the signal wavelength (usually $\delta=0.5$). Assuming line-of-sight signal propagation and a single-antenna transmitter, the channel between the latter and the RIS is given by
\begin{align}
    \g \triangleq \sqrt{\gamma_t}\, \a(\theta_t,\phi_t) \in \Compl^{N\times 1} ,
\end{align}
where we define the average channel power gain as $\gamma_t \triangleq {\beta_0}/{d_t^2}$, with $\beta_0$ the average channel power gain at a reference distance. $d_t$ represents the distance between the transmitter and the RIS, whereas $\theta_t$, $\phi_t$ denote the angles of arrival at the RIS along the azimuth and elevation, respectively. With similar reasoning, the channel between the RIS and the single-antenna receiver is given by
\begin{align}
    \h \triangleq \sqrt{\gamma_r}\, \a(\theta_r,\phi_r) \in \Compl^{N\times 1}.
\end{align}
The matrix containing the RIS configuration is defined as
\begin{align}
    \Phib \triangleq \mathrm{diag}[e^{j\psi_1},\ldots,e^{j\psi_N}] \in \Compl^{N\times N} , \label{eq:Phi_RIS}
\end{align}
with $\psi_n \in \mathcal{Q}, \, \forall n$ and $\mathcal{Q}$ the quantized RIS phase shift set. Note that, to preserve a tractable model in \eqref{eq:Phi_RIS}, we have neglected any phase-dependent reflection coefficient at the RIS. Lastly, the received signal at the receiver is given by
\begin{align}
    y \triangleq \h^\herm \Phib \g \, s + n \in \Compl ,
\end{align}
where $s\in \Compl$ is the transmitted symbol and $n\in \Compl$ is the noise term distributed as $\mathcal{CN}(0,\sigma_n^2)$. 

Let $\v \!\triangleq \!\mathrm{diag}(\Phib^\herm)\!\in\!\Compl^{N\times 1}$ and $\bar{\h} \!\triangleq \!\mathrm{diag}(\h^\herm)\g \!\in\!\Compl^{N\times 1}$, such that the power at the receiver is maximized by letting \cite{ZhangActivePassive}
\begin{align}
    \v = \mathrm{exp}[j \, f_q(\angle \bar{\h})],\label{eq:v_opt}
\end{align}
where $f_q(\cdot)$ projects each element of the vector $\angle\bar{\h}$ in \eqref{eq:v_opt} onto the closest element of set $\mathcal{Q}$ to obtain a feasible solution. 

It is important to highlight that vector $\bar{\h}$ has the form of a scaling term times the UPA response vector in \eqref{eq:UPA} for some steering angles $({\theta},{\phi})$. Hence, in order to design a codebook of RIS beamforming vectors, we artificially create $N_{\mathrm{B}}$ pairs of $\{({\theta_n},{\phi_n})\}_{n=1}^{N_{\mathrm{B}}}$ couples and generate the corresponding UPA response vectors $\{\bar{\h}_n\}_{n=1}^{N_{\mathrm{B}}}$. Given the symmetry of the array response around the $x$-axis for the azimuth and around the $y$-axis for the elevation, and in the interest of saving measurement time, we sample a regular grid of points spaced by $3$ degrees in the search space $[-\pi/2,\pi/2]\times[-\frac{\pi}{4},\frac{\pi}{4}]$, such that $N_{\mathrm{B}} = 1891$. Finally, we obtain the RIS configurations $\mathcal{V} := \{\v_n\}_{n=1}^{N_{\mathrm{B}}}$ for each steering angle couple $\{({\theta_n},{\phi_n})\}_{n=1}^{N_{\mathrm{B}}}$ in the codebook by applying the expression in \eqref{eq:v_opt}.



\section{Prototype Implementation}\label{sec:prototype}
We prototyped our design using a two-layer PCB (Printed Circuit Board). 
%
%
The substrate material is FR-4, a composite material made of woven fiberglass with an epoxy resin binder that is flame resistant. Its relative electrical permittivity is in the range of  $\epsilon_r = [4.1, 4.8]$, and is coated by two layers of 1-ounce copper (35$~\mu \text{m}$). 
In general, thick substrates and high permittivity lead to small bandwidths and low efficiency due to surface waves~\cite{pandey2019practical}. 
Since the operating frequency of our prototype is $f=5.3$~GHz ($\lambda=56.56$~mm), we chose a substrate thickness of $h=0.53$~mm, that is in the range $0.003 \lambda \leq h \leq 0.05 \lambda$ as suggested in~\cite{pandey2019practical}.

\begin{figure}[b!]
\vspace{2mm}
\minipage{0.49\columnwidth}
\centering
    \includegraphics[width=0.75\columnwidth]{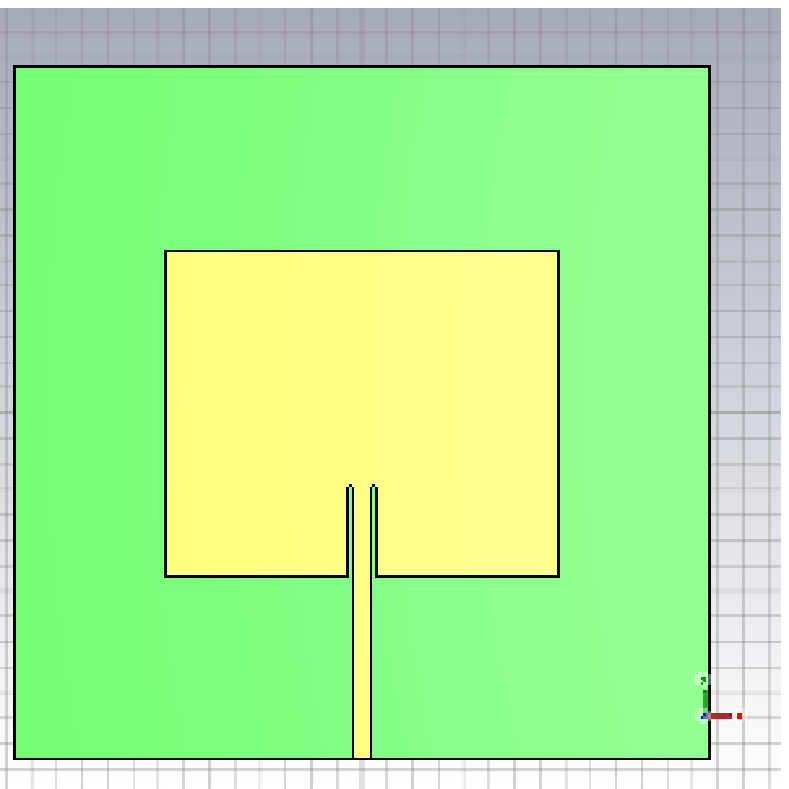}
    \vspace{-4mm}
\caption{\small Layout of the patch antenna design.}
    \label{fig:patch}
\endminipage
\hfill
\minipage{0.47\columnwidth}
    \includegraphics[width=\columnwidth]{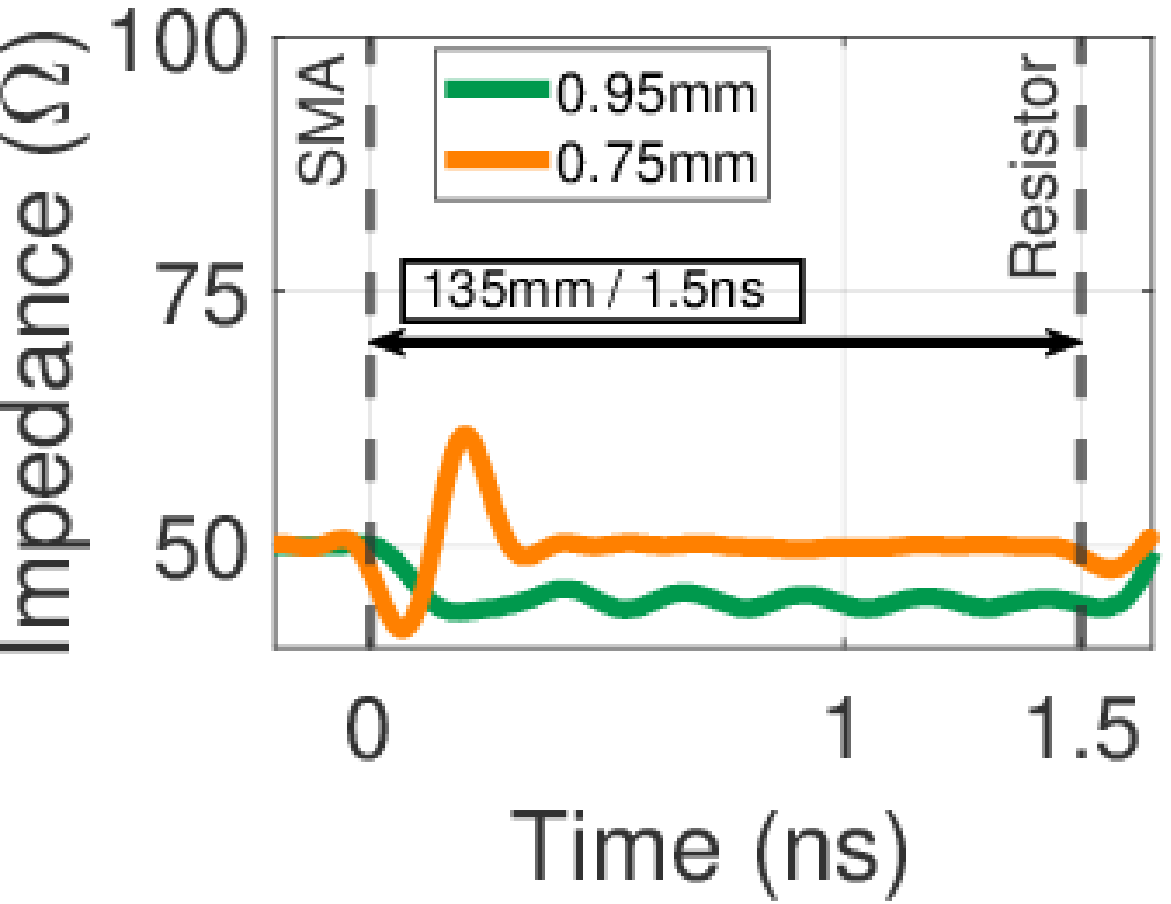}
      \vspace{-8mm}
\caption{\small TDR chart of a 0.95mm-width microstrip.}
    \label{fig:TDR}
\endminipage

\end{figure}

\subsection{Patch Antenna} \label{sec:proto:antenna}
Patch antennas are implemented by cutting out a particular shape from the copper of the board's upper layer. In this way, the remaining metallic shape can radiate at the desired frequency while the back layer operates as ground for the antenna. 
%
Following the conventional literature on antenna design, we used a rectangular shape, as shown in Fig.~\ref{fig:patch}. In more detail, we used the transmission-line model from \cite{balanis} to calculate its width $W$ and length $L$ as follows:
\begin{align} 
    W &= \frac{\lambda}{2 \sqrt{0.5(\epsilon_r+1)}} = 16.9~\text{mm}, \label{eq:W} \\
    L &= L_{eff} - 2\Delta L = 13.15~\text{mm}, \label{eq:L}
\end{align}
 where $\epsilon_{eff} = \frac{\epsilon_r + 1}{2} + \frac{\epsilon_r - 1}{2} (\frac{1}{\sqrt{1 +12 \frac{h}{w}}})$ is the effective dielectric constant that takes into account the fact that the electric field lines reside in the substrate and partially in the air, $L_{eff} = \frac{c}{2 f \sqrt{\epsilon_{eff}}}$ is the effective length, $c$ is the speed of light, and $\Delta L = 0.412 h \cdot \frac{\epsilon_{eff} + 0.3}{\epsilon_{eff} - 0.258} \cdot \frac{\frac{h}{w} + 0.264}{\frac{h}{w}+0.8}$ is an offset to obtain the antenna physical length from $L_{eff}$ (see \cite{balanis} for details).

As shown in Fig.~\ref{fig:patch}, a microstrip connects each antenna to the RF switch. To this end, we selected an \emph{inset feeding} approach, with a notch at the edge of the antenna. This approach allows us to adapt the antenna to a precise characteristic impedance, which is crucial to maximizing power transfer. Given this notch (see details later), we refined the geometry derived before with the parameters shown in Fig.~\ref{fig:patch} by exhaustive search using a full-wave simulator~\cite{cststudio}, thus setting $W = 15.5~\text{mm}$ and $L = 12.8~\text{mm}$. 
%

In the following we describe in detail two crucial parameters, namely ($i$) the width of the microstrip that connects the RF switch (see Fig.~\ref{fig:unit_cell}), and ($ii$) the position of the notch. The former is essential to guarantee impedance-matching, and, since the usual characteristic impedance is 50~$\Omega$, the width of all the microstrips must be selected accordingly. 

\subsubsection{Width of the feeding line}\label{sec:proto:width_lines}

To avoid power loss between the feeding line and the RF switch, their  characteristic impedance should match. 
Therefore, we used the model described in \cite{strip} to estimate a $0.95$-mm line width, which equalizes the $50\Omega$-impedance of the switch. 

%
To validate this, we printed a $0.95$mm-width microstrip and applied the Time Domain Reflectometry (TDR) technique using a Vector Network Analyzer (VNA)  to measure the actual characteristic impedance along the line. TDR generates a pulse with a short rising time that allows us to calculate the impedance along the line based on the received reflections.


As depicted in Fig.~\ref{fig:TDR} (green), this experiment shows that the actual impedance along the line is smaller than the predicted $50~\Omega$. Such a mismatch with the impedance of the switch would incur some power loss at every unit cell and, consequently, poor beamforming gains overall.
%

\begin{figure}[t!]
\minipage{0.49\columnwidth}
\centering
  \includegraphics[width=0.75\linewidth]{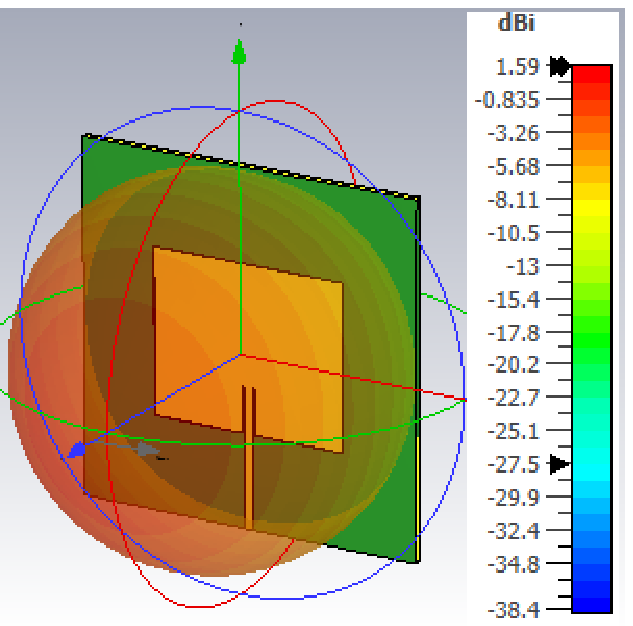}
  \vspace{-4mm}
  \caption{\small Expected radiation pattern of the patch antenna.}
  \label{fig:far_f}
\endminipage
\hfill
\minipage{0.49\columnwidth}
  \includegraphics[width=\linewidth]{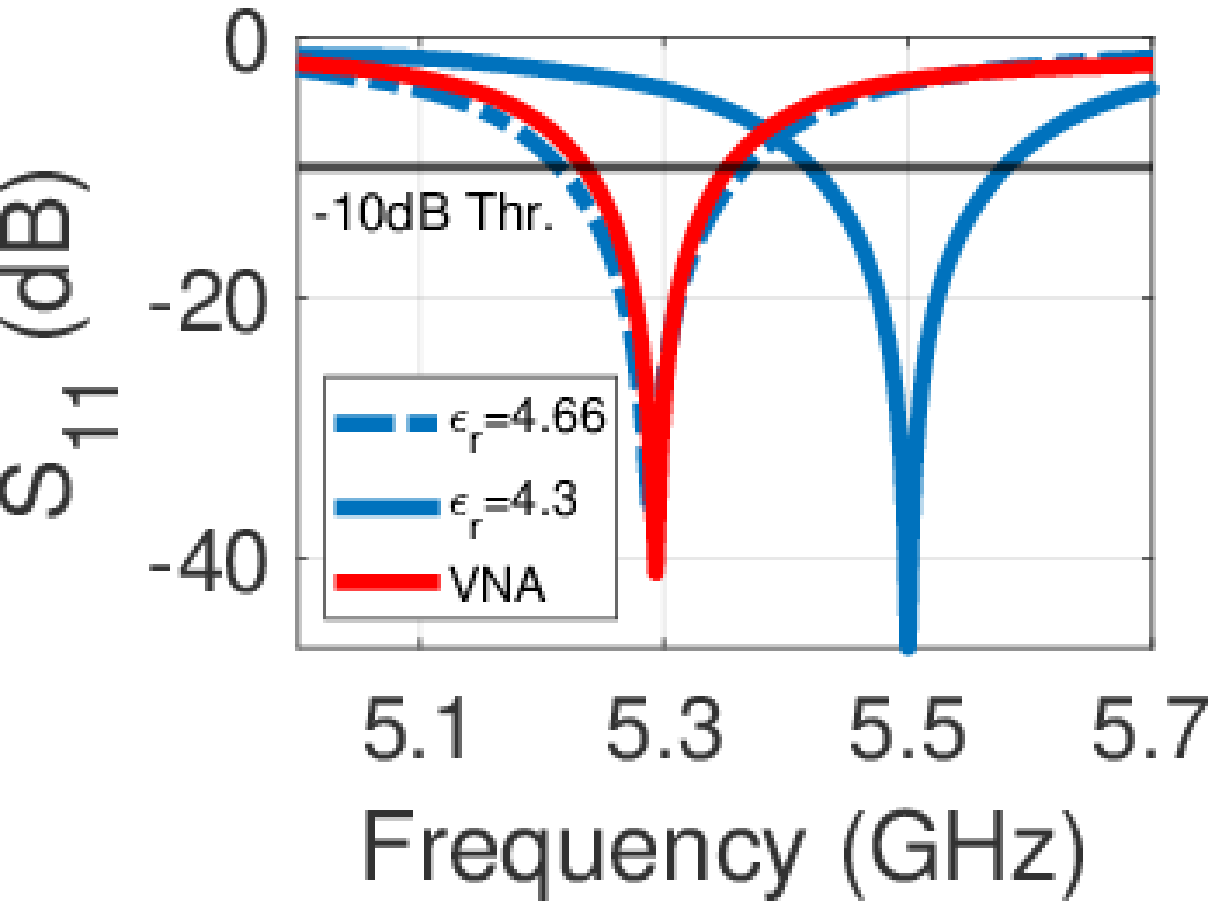}
    \vspace{-8mm}
  \caption{\small $\boldsymbol{S_{11}}$ parameter of two patch antenna samples.}
  \label{fig:s11}
\endminipage
\end{figure}

Consequently, we opted for a simple empirical approach: we printed out several $135$mm-length microstrips with different widths, and applied the TDR method to each sample.
Fig.~\ref{fig:TDR} shows with an orange line the result of the selected sample, with a width equal to 0.75~mm, which provided the best performance.  Ignoring the large oscillation at the beginning of the line, which is due to the soldered SMA connector that we used to connect the line and the VNA, the experiment shows a perfect match with the expected value of $50~\Omega$. 

\subsubsection{Notch} 
The second relevant parameter is the depth of the notch, which should minimize the amount of reflected power. This is achieved when the impedance of the antenna matches that of the feeding line. However, the antenna's impedance diminishes as one moves towards its center because the current's intensity is higher at that point. 
Hence, it is important to carefully design the depth of the notch.

Following the analytical method introduced in \cite{balanis}, we first estimated the impedance at the bottom edge of the antenna and centered on the horizontal plane (see red bullet in Fig.~\ref{fig:patch}). At this point, the impedance is purely resistive, i.e., its reactance is zero, and should be equal to  $R_{edge}\approx 341~\Omega$ (we omit the details of the mathematical model, which can be found in \cite{balanis}, to reduce clutter). 
Then, the optimal depth of the notch can be computed as:
\begin{align}
    h_{\text{notch}} = \frac{L}{\pi} \cos^{-1}\left( \sqrt{\frac{R}{R_{edge}}}\right) \approx 4.9~\text{mm},
\end{align}
where $R$ is the desired impedance (i.e., $50~\Omega$). 

We attempted to validate this result with the full-wave simulator~\cite{cststudio} and found that a $3.5$-mm depth, cutting across the patch antenna as shown in Fig.~\ref{fig:patch}, maximizes performance ($\sim\!\!30\%$ difference with respect to the model). Fig.~\ref{fig:s11} shows with a blue line the amount of power that is reflected, estimated by the simulator and referred to as S11 parameter in antenna design. The result shows good performance at 5.5~GHz, the operating frequency of choice.
%
%
Fig.~\ref{fig:far_f} shows the simulated radiation pattern, with minimal backwards propagation.  Note that the antenna is not very directive and has an expected gain of 1.5dBi, which is common for this type of low-cost antennas.  

To validate the patch design, we printed a sample antenna with the aforementioned parameters. Using our VNA, we measured the empirical $S_{11}$ and plotted the result with a red line in Fig.~\ref{fig:s11}. Perhaps surprisingly, the minimal-$S_{11}$ frequency point is 5.3~GHz instead of the intended 5.5~GHz. After some research, we realized that the offset stems from an error on the nominal permittivity $\epsilon_r=4.3$ used in our model/simulations for the PCB substrate. After some iterations with our simulator, we estimated the real permittivity to be $\epsilon_r=4.66$. In light of this, we changed the operating frequency to $f=5.3$~GHz. 

From Fig.~\ref{fig:s11}, we can also estimate that the bandwidth of our approach, i.e., the range of frequencies where the antenna's $S_{11}$ is $\le-10$~dB, is 118MHz (note the black horizontal line).  We finally note  that, at 5.3~GHz, the $S_{11}$ is $-41$~dB, which corresponds to a Voltage Standing Wave Ration (VSWR) of $1.018$. This means that the amount of power from the feeding line that is reflected back is negligible, which was our goal.

\begin{figure}[t!]
\centering
\includegraphics[width=0.6\columnwidth]{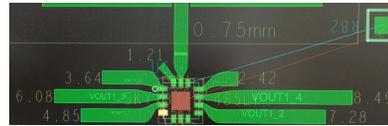}
\vspace{-4mm}
\caption{\small RF switch and microstrips.}
\label{fig:RFconnection}
\end{figure}

\setlength\tabcolsep{1.5pt} 
\begin{table}[t!]
\vspace{-3mm}
\centering
\footnotesize
\begin{tabular}{|r|c|c|c|c|c|c|c|}
\hline
\textbf{Output port}         & 7     & 6      & 5      & 1      & 3      & 2      & 4    \\ \hline
\textbf{$\varphi$ {(}deg{)}} & 51.42 & 102.85 & 154.28 & 205.71 & 257.14 & 308.57 & 360  \\ \hline
\textbf{$l$   {(}mm{)}}       & 1.21  & 2.42   & 3.64   & 4.85   & 6.08   & 7.28   & 8.49 \\ \hline
\end{tabular}
\caption{\small RF switch's output ports, the length of the associated delay lines, and the resulting phase shifts.}
\label{tab:3bitphase}
\vspace{-3mm}
\end{table}

\subsection{Phase shifters}

At each cell, a specific phase shift $\varphi$ is applied by \emph{routing} the RF signal towards a specific delay line, implemented with a microstrip, that reflects the signal back to the patch antenna. To this end, we use a 3-bit RF switch SKY13418-485LF \cite{SKY}, which has 1 input port (attached to the feeding line), 3 configuration ports (more later), and 8 output ports connected to delay lines of different lengths. Given one configuration (input-output port mapping encoded as 3 bits in the configuration ports), the resulting phase shift follows as:
\begin{equation}
\varphi = \frac{360 \cdot 2l \cdot f}{c \cdot v_f},
\label{phase}
\end{equation}
where $l$ denotes the distance travelled from the patch antenna to the end of the delay line (including the switch and the feeding and delay lines), and $v_f$ is the velocity factor of the microstrip material. 
Note that $2l$ accounts for the round-trip between antenna and delay line. 

To estimate $v_f$ empirically, we take advantage of the TDR technique used earlier, which also measures the time $d_{\mu \text{strip}}$ it takes for a signal to travel through a microstrip of length $l_{\mu \text{strip}}$. As shown in Fig.~\ref{fig:TDR}, $d_{\mu \text{strip}}= 1.51$~ns for a line of $l_{\mu \text{strip}}=135$~mm, which is sufficiently long to force the signal to travel at least $2\lambda$ and hence enable highly accurate delay estimates. We then calculate $v_f = \frac{v_{\mu \text{strip}}}{c} = 0.298 \approx 0.3$, where $v_{\mu \text{strip}}=\frac{l_{\mu \text{strip}}}{d_{\mu \text{strip}}}$ is the velocity of the signal through the microstrip.

Given $v_f$ and the selected microstrips width derived in \S\ref{sec:proto:width_lines} ($0.75$~mm), we can calculate the length of the delay lines corresponding to the phase shifts that need to be encoded into each output port, as indicated in Table~\ref{tab:3bitphase}. Note that port 8 has no associated phase shift. Instead, this port connects with a delay line that ends with a $50\Omega$-resistor, which prevents the signal to be reflected back. We refer to this configuration as ``absorption state'', which enables us to build \emph{virtual} RISs of any size and shape. Alternatively, an energy harvester~\cite{mir2021passivelifi} can be used to feed the MCU and effectively make it self-sustainable. The resulting design is shown in  Fig. \ref{fig:RFconnection}.

\subsection{Microcontroller Unit (MCU)}

As explained in \S\ref{sec:design}, an MCU is in charge of parametrizing the configuration ports of the RF switch in each unit cell.
We have selected the STM32L071V8T6 MCU from STMicroelectronics~\cite{mcu}, which is low cost (see \S\ref{sec:cost}), high-speed (configuring 100 cells takes $<35$~ms), and low energy-consuming ($62$~mW in high-performing mode). 
We note that we have not optimized the MCU, which we leave for future work. For instance, we only use the MCU's high-performance (high-consuming) mode, although it provides low-consuming modes too. Exploiting these modes could reduce its energy consumption to the order of $\mu$W, which is amenable to energy harvesters~\cite{mir2021passivelifi}. 

\begin{figure}[t!]
     \centering
     \begin{subfigure}[b]{0.49\linewidth}
         \centering
         \includegraphics[width=0.93\textwidth]{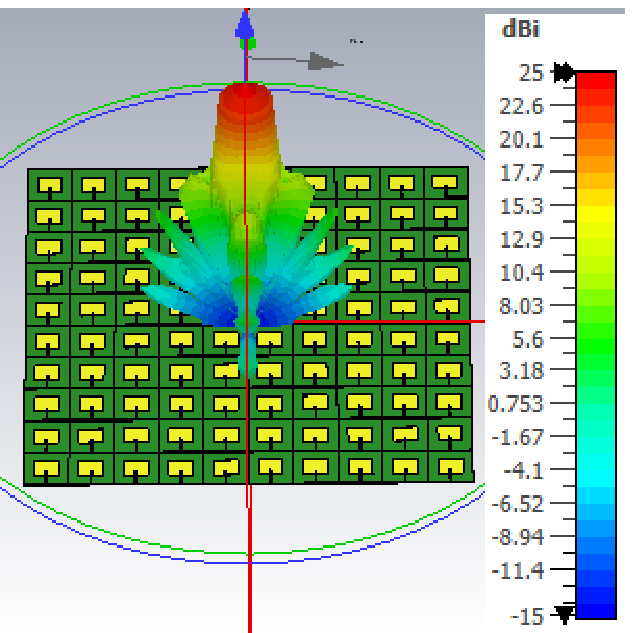}
         \caption{$\boldsymbol{\theta=0^\circ, \phi=0^\circ}$.}
         \label{fig:ris_0azimuth}
     \end{subfigure}
     \hfill
     \begin{subfigure}[b]{0.49\linewidth}
         \centering
         \includegraphics[width=0.93\linewidth]{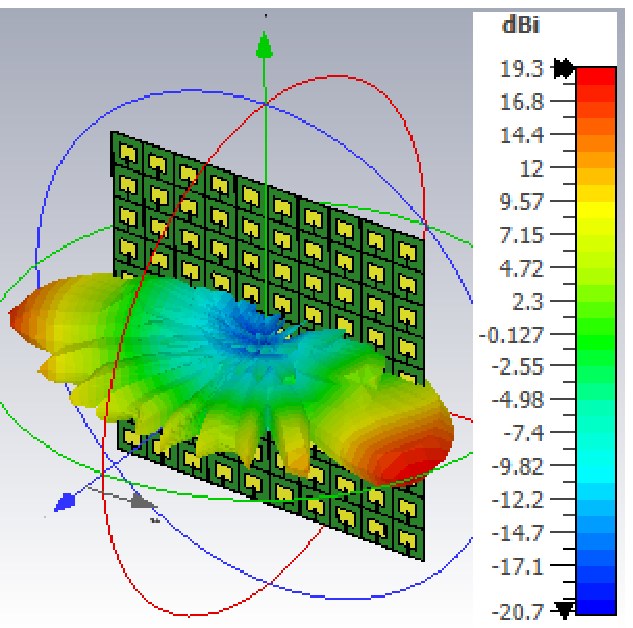}
         \caption{$\boldsymbol{\theta=80^\circ, \phi=0}$.}
         \label{fig:ris_80azimuth}
     \end{subfigure}
     \vspace{-4mm}
        \caption{\small Expected beamforming pattern of a 10$\boldsymbol{\times}$10 RIS.}
        \label{fig:ris_azimuth}
\end{figure}

\subsection{RIS board}\label{sec:proto:board}

The spacing between unit cells (antennas) has to be carefully designed to maximize beamforming gains. Roughly speaking, a small spacing increases the probability of mutual coupling, which decreases the efficiency of each antenna because of surface waves propagation. Conversely, a large spacing leads to grating lobes, as we demonstrate in \S\ref{sec:results}. All in all, the inter-cell spacing depends on the maximum steering angle $\theta_{max}$ of the main lobe, which is given by
$
d_{max} = \frac{\lambda}{1+\sin(\theta_{max})} \label{eq:steering}
$.
Note that, though $d_{max} =\lambda/2$ maximizes the steering angle range of the array, $\theta_{max}=90^{\circ}$ is not achievable in practice~\citep{AD}.

Prior to developing a prototype, we simulated a $10\times10$ RIS board with a regular $10\times10$ planar antenna array. 
%
In our first set of simulations, each antenna element is fed with equal power from an open-ended transmission line that induces a phase delay that is optimized offline to maximize power towards the selected azimuth and elevation angles.  
%
Figs.~\ref{fig:ris_0azimuth} and \ref{fig:ris_80azimuth} show the expected radiation pattern of two different configurations that maximize power towards an elevation of $\phi=0^\circ$ and, respectively, an azimuth of $\theta=\{0^\circ, 80^\circ\}$. 

On the one hand, we can observe from Fig.~\ref{fig:ris_0azimuth}  that the array can achieve a narrow beampattern, with a Half Power Beamwidth  (HPBW) equal to $10.1^\circ$, with a gain of $25$~dBi in the intended direction, which halves with a 10-degree offset. No back-radiation is expected. On the other hand, Fig.~\ref{fig:ris_80azimuth} shows that a large steering angle of $\theta=80^\circ$  dissipates half the energy towards the opposite direction and drops the power of the beam to $19.3$~dBi. 

\begin{figure}[t!]
     \centering
     \begin{subfigure}[b]{0.44\linewidth}
         \centering
         \includegraphics[width=0.75\columnwidth]{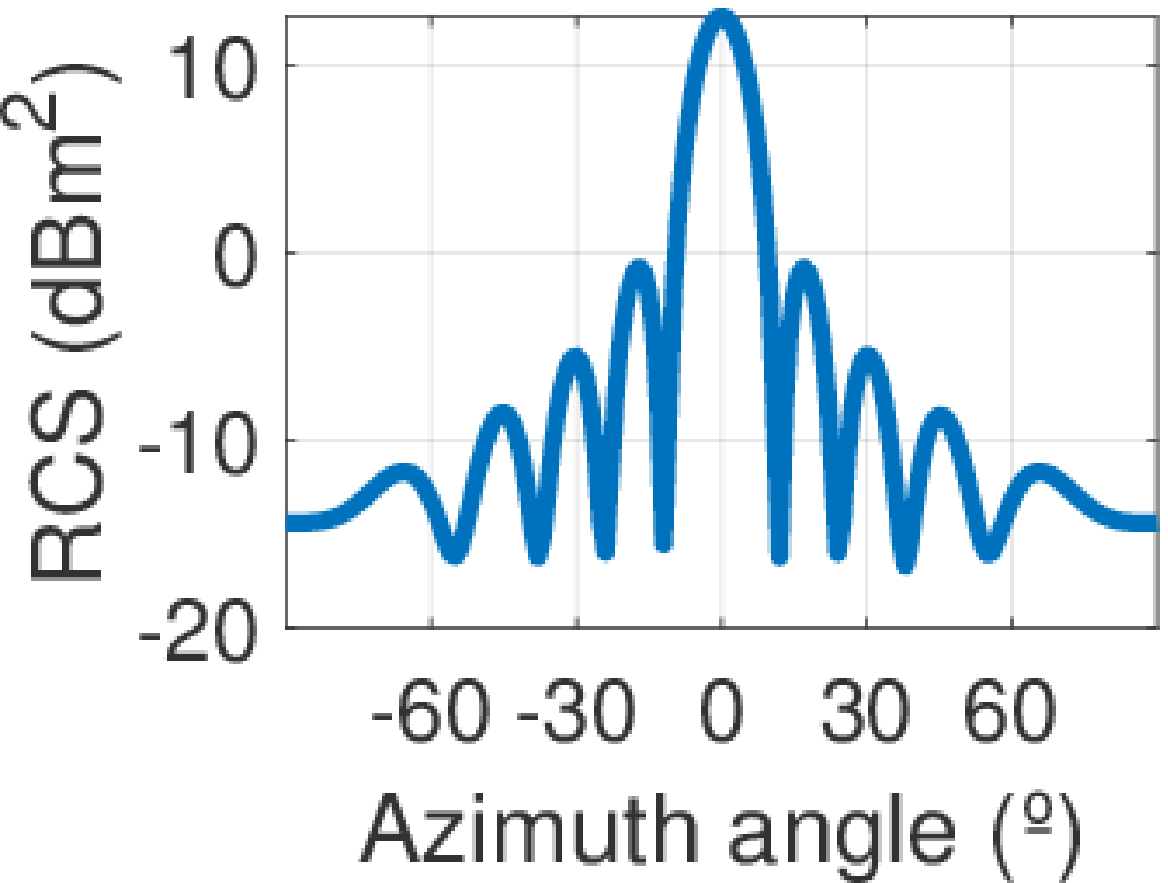}
         \vspace{-2mm}
         \caption{$\boldsymbol{0^\circ}$ incidence angle.}
         \label{fig:rcs0}
     \end{subfigure}
     \hfill
     \begin{subfigure}[b]{0.44\linewidth}
         \centering
         \includegraphics[width=0.75\linewidth]{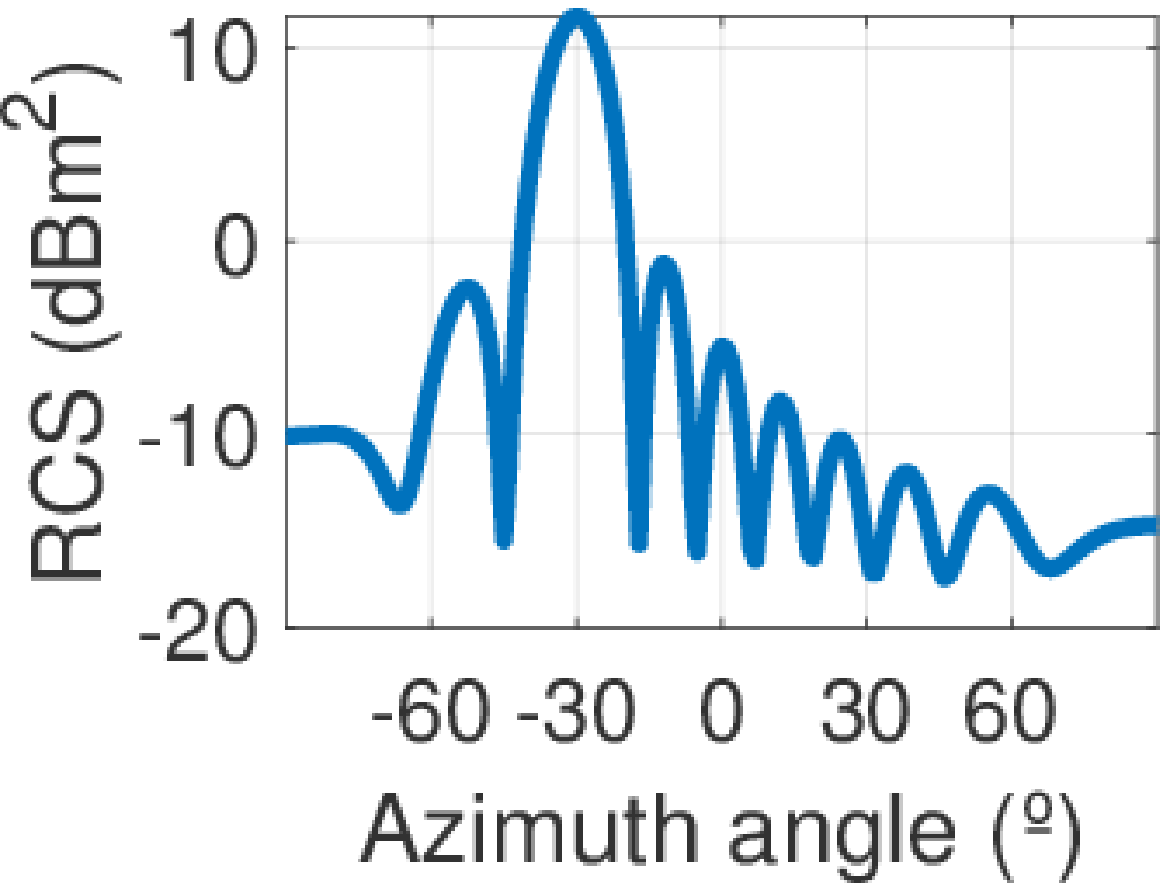}
         \vspace{-2mm}
         \caption{$\boldsymbol{30^\circ}$ incidence angle.}
         \label{fig:rcs30}
     \end{subfigure}
        \vspace{-4mm}
        \caption{\small RCS of a 10$\boldsymbol{\times}$10 RIS illuminated by a plane wave.}
        \label{fig:rcs}
\end{figure}

\begin{figure}[t!]
    \vspace{-4mm}
    \centering
    \begin{subfigure}[b]{0.49\linewidth}
        \centering
        \includegraphics[width=\columnwidth]{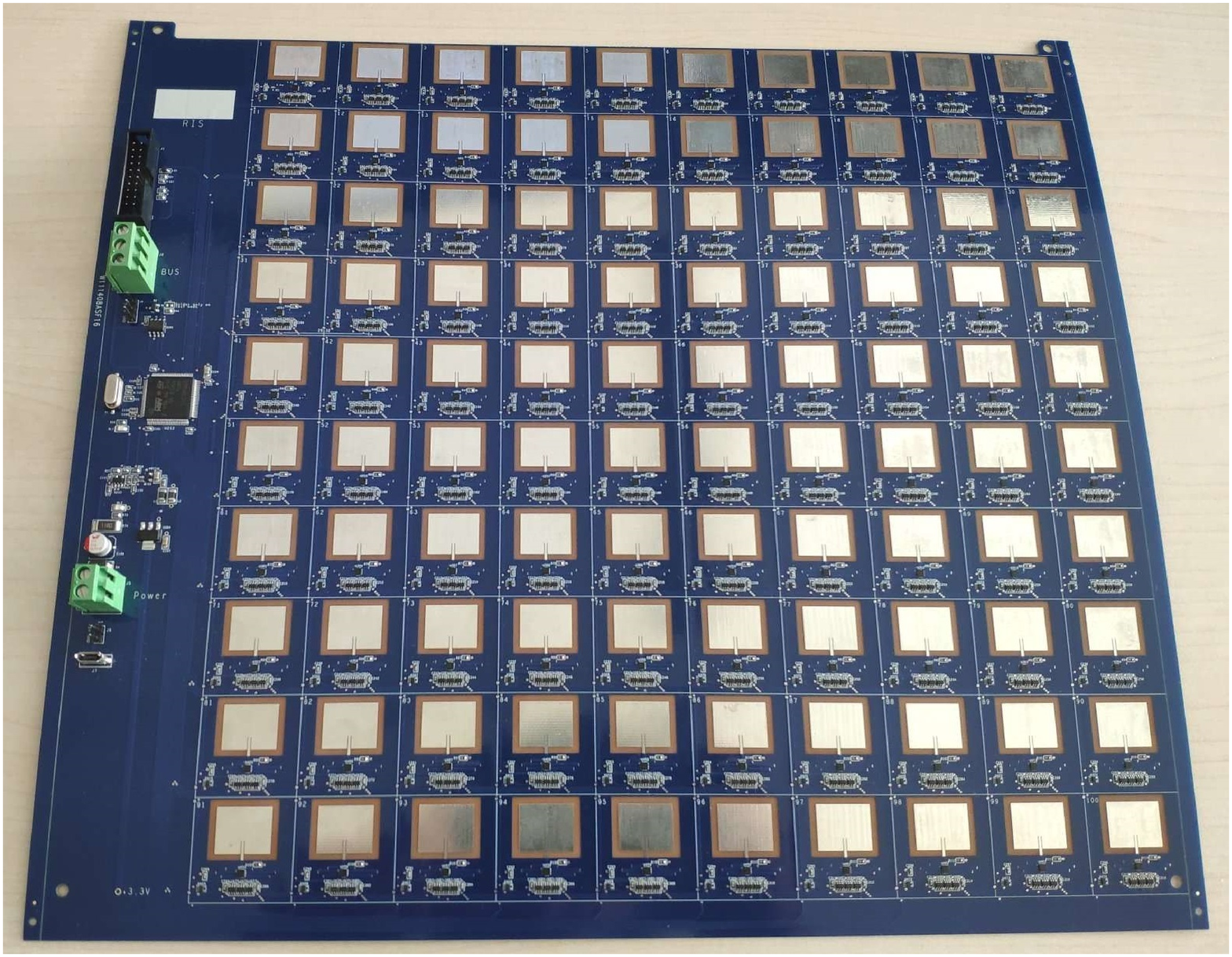}
        \caption{RIS board.}
        \label{fig:ris_pcb}        
    \end{subfigure}
    \hfill
    \begin{subfigure}[b]{0.49\linewidth}
        \centering
        \includegraphics[width=\textwidth]{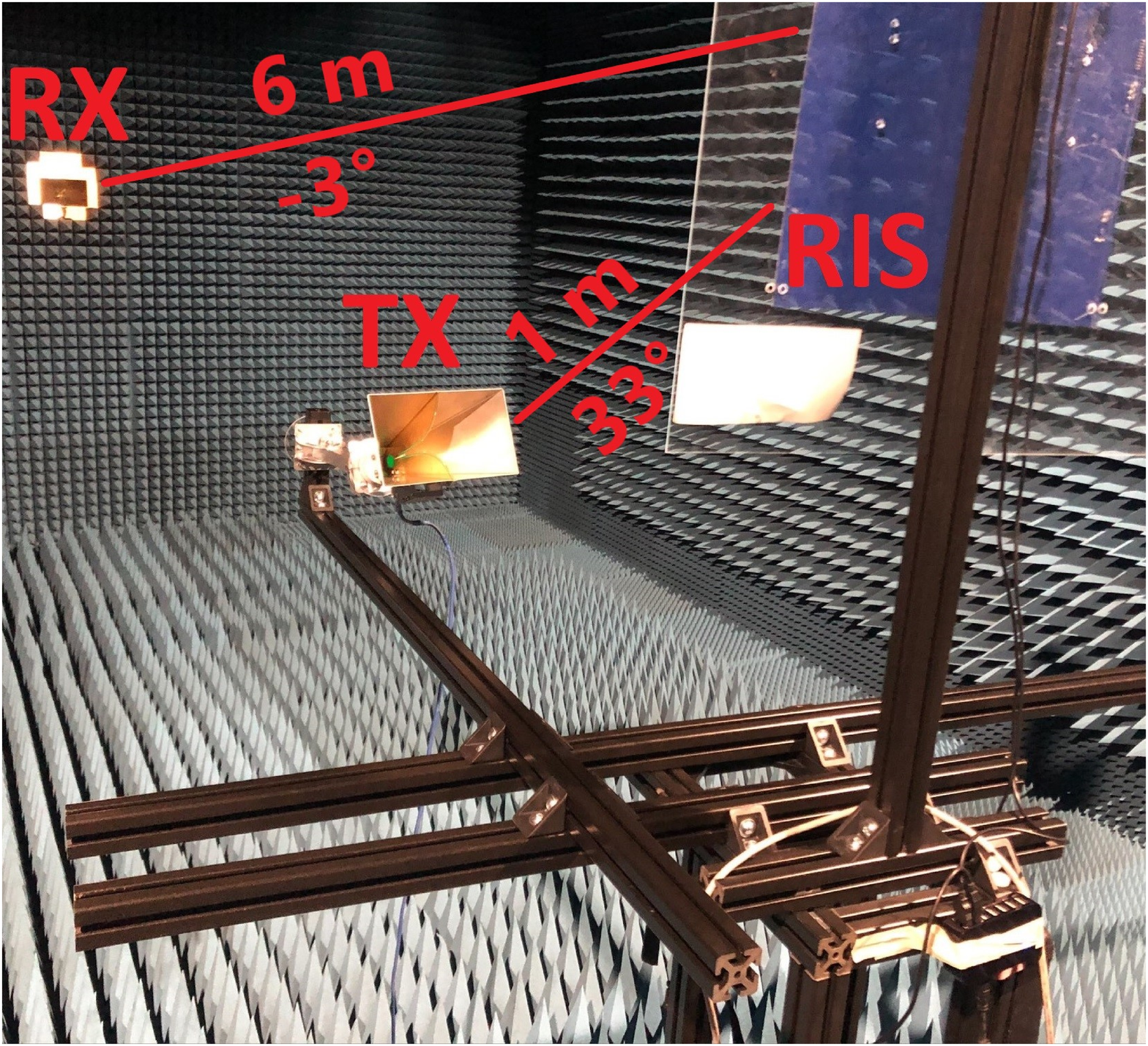}
        \caption{Testbed.}
        \label{fig:ris_anech}        
    \end{subfigure}
    \vspace{-4mm}
    \caption{\small (a) 10$\boldsymbol{\times}$10 RIS PCB realization, and (b) testbed in an anechoic chamber with rotating structure.}
\end{figure}

Note however that the power radiated by a RIS comes from an external source that \emph{illuminates} the surface. In order to simulate this behavior, we generated a plane wave linearly polarized, and measured the resulting Radar Cross Section (RCS). The RCS estimates how much of the incident power at every point of the surface is scattered back to the receiver. Hence, we expect high RCS values in the direction of the main beam and lower in other directions. This is indeed confirmed in Figs.~\ref{fig:rcs0} and \ref{fig:rcs30} for an incidence angle of $0^\circ$ and $30^\circ$, respectively. In both cases, the RIS is configured to reflect the received signal perpendicularly to the incidence angle of the received signal, i.e., the RIS should maximize power in the same direction of the incidence angle, which is confirmed by both figures with a RCS approximately equal to $12-13$~dBsm in the intended direction.  This highlights the fact that, in real life scenarios, the angle of arrival must be known to the controller to maximize beamforming gains.


We manufactured 10 boards of 10x10 unit cells each. All the phase and selection buses, which connect each unit cell with an MCU, are built with microstrips. Unit cells are deployed in the PCB layout such that the same inter-cell distance can be maintained across co-located boards. The remaining elements described in \S\ref{sec:design} (flip-flops, resistors, etc.), are standard components assembled on the PCB. The final printout is shown in Fig.~\ref{fig:ris_pcb}.

\begin{figure}[t!]
    \centering
    \includegraphics[width=0.7\columnwidth]{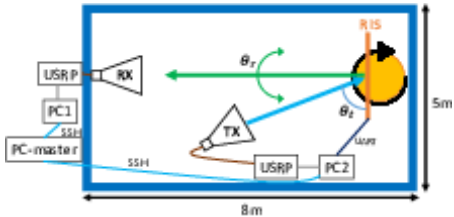}
    \vspace{-4mm}
    \caption{\small Map of the testbed.}
    \label{fig:testbed}
\end{figure}

\vspace{-3mm}
\section{Empirical Characterization}\label{sec:results}

We characterized one of our 10x10 RIS boards in an 8m$\times$5m anechoic chamber. Figs.~\ref{fig:ris_anech} and \ref{fig:testbed} illustrate our testbed. 
We mounted the board on a turntable controlled remotely from a master PC, which is also used to configure the beamforming parameters of the RIS. 
We use two software-defined radio devices attached to horn antennas with gain $G=13.5$~dBi to generate (``TX'') and receive (``RX'') a continuous stream of OFDM QPSK-modulated symbols with 5~MHz of bandwidth and numerology that meets 3GPP LTE requirements. 
%
The transmission power of TX is -30~dBm per subcarrier, and we sample the reference signal received power (RSRP) at RX. 

The distances RIS-TX and RIS-RX are $d_{\text{RIS}-\text{TX}}\!=\!1.1$~m and $d_{\text{RIS}-\text{RX}}\!=\! 6.3$~m, respectively. Considering the size of the RIS and its operating frequency, it is hard to guarantee that $d_{\text{RIS}-\text{TX}}$ is larger than the far-field threshold, which is $2\frac{D^2}{\lambda}\!=\!6.5$~m~\cite{balanis}, where $D\!=\!0.43$~m is the diagonal of the array. Nevertheless, our choice of $d_{\text{RIS}-\text{TX}}$ is larger than the \emph{reactive} near-field threshold, which is $0.62\sqrt{\frac{D^3}{\lambda}}\!=\!0.73$~m~\cite{balanis}, and sufficient for our purposes.
As shown in Fig.~\ref{fig:testbed}, the rotation of the table determines the azimuth angle $\theta_r$, and the location of TX determines $\theta_t$. Conversely, the elevation angles of RIS-TX and RIS-RX are fixed to $\phi_t \!=\! 33^{\circ}$ and $\phi_r \!=\! -3^\circ$, respectively.

\vspace{-1mm}
\subsection{Codebook characterization}

\begin{figure}[t!]
     \centering
     \includegraphics[width=\columnwidth]{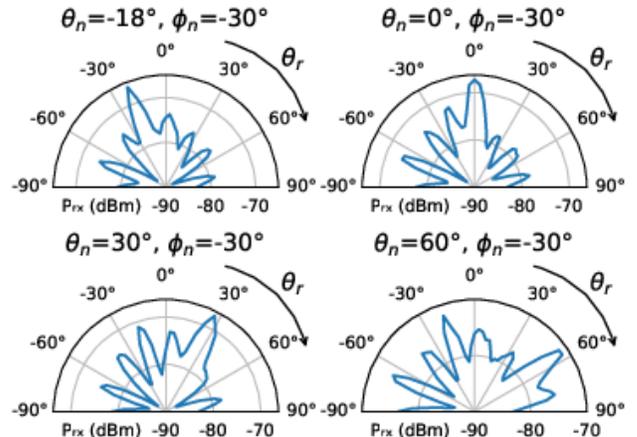}
    \vspace{-14mm}
     \caption{Examples of beampatterns for $\boldsymbol{\theta_t=0^\circ}$.}
     \label{fig:pattern:1}
\end{figure} 

We begin our experimental campaign by characterizing the codebook generated in \S\ref{sec:model}. 
To this end, we test out all the configurations $\mathcal{V} = \{\v_n\}_{n=1}^{N_{\mathrm{B}}=1891}$ for a wide range of $\theta_r=[-90^{\circ}, 90^{\circ}]$ and for $\theta_t=\{20^{\circ}, 90^{\circ}\}$.

\begin{figure}[t!]
     \centering 
     \includegraphics[width=\columnwidth]{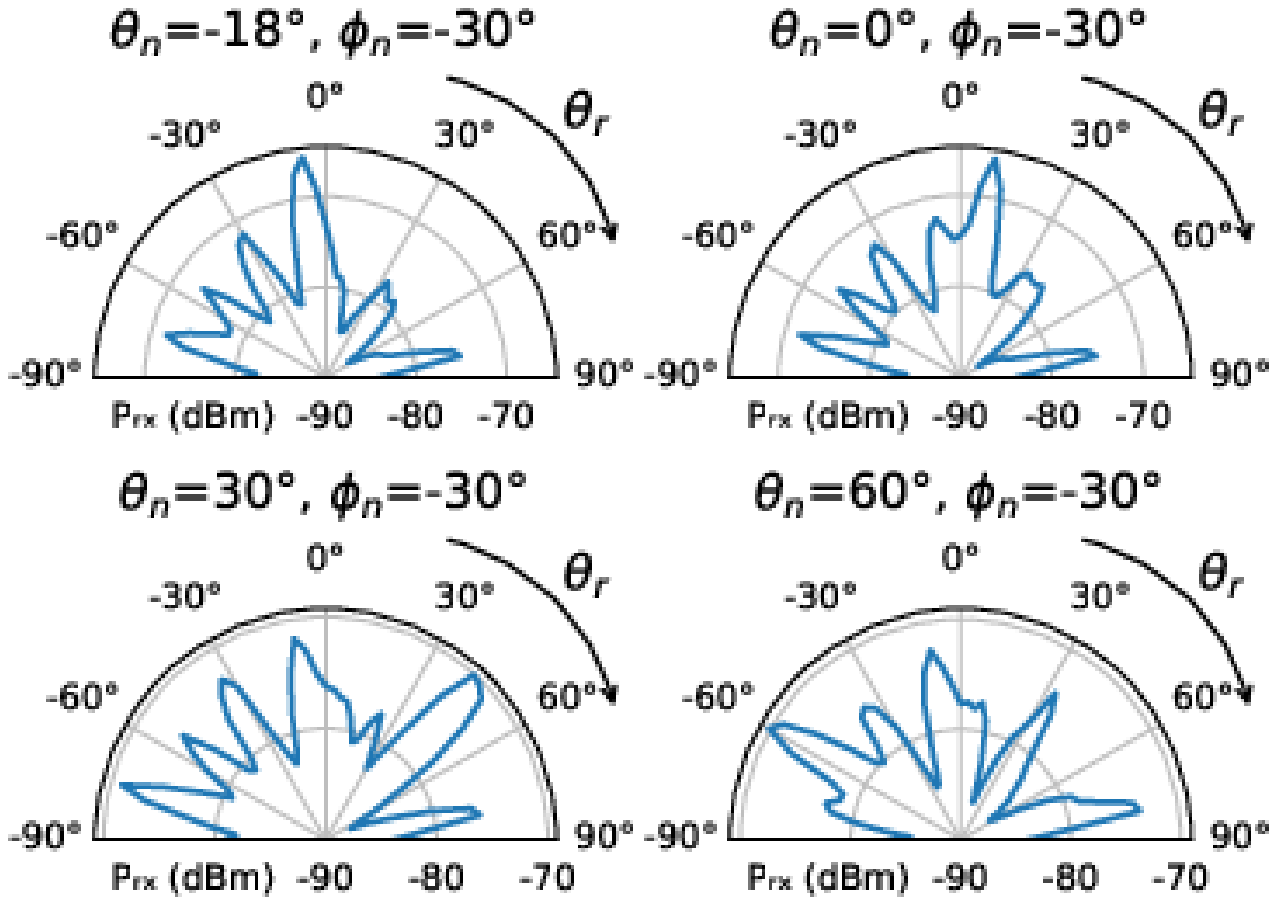}
    \vspace{-14mm}
     \caption{Examples of beampatterns for $\boldsymbol{\theta_t=20^\circ}$.}
     \label{fig:pattern:2}
\end{figure}

For our empirical results, the first observation is that the direction $(\theta_n, \phi_n)$ of the main lobe points towards $(\theta_r- \theta_t, \phi_r- \phi_t)$, as intended, for all configurations $\v_n\in\mathcal{V}$. These results, hence, validate our prototype for practically all configurations in $\mathcal{V}$. 
Figs.~\ref{fig:pattern:1} and \ref{fig:pattern:2} depict some representative configurations $\v_n\in\mathcal{V}$ for both $\theta_t$ settings, respectively. These figures show that the main beam points towards the intended directions.
%
%
%
%
We note, however, that the gain of the main lobe is penalized when we use large steering angles (see, e.g., $\theta=60^{\circ}$ in both figures), which is expected~\cite{AD}. 
%
Overall, the power received in the intended direction ranges between $-74$~dBm (for large steering angles) and $-64$~dBm (for smaller angles), which give us remarkable beamforming gains between $\sim\!17$~dB and $\sim\!27$~dB over the noise floor. 

Using the radar range equation in \cite{balanis}, the peak RCS can be calculated as
$
64\pi^3 \frac{P_{RX}}{P_{TX}} \cdot (\frac{d_{\text{RIS}-\text{TX}} \cdot d_{\text{RIS}-\text{RX}}}{\lambda \cdot G})^2
= 11.2$~dBm$^2$. Moreover, the HPBW is in average around $10^\circ$ for all beampatterns. Both results are in line with our simulations in \S\ref{sec:proto:board}. 


\vspace{-1mm}
\subsection{Scalability analysis}

\begin{figure}[t!]
\vspace{-4mm}
     \centering
     \begin{subfigure}[b]{0.15\columnwidth}
         \centering
         \includegraphics[width=\textwidth]{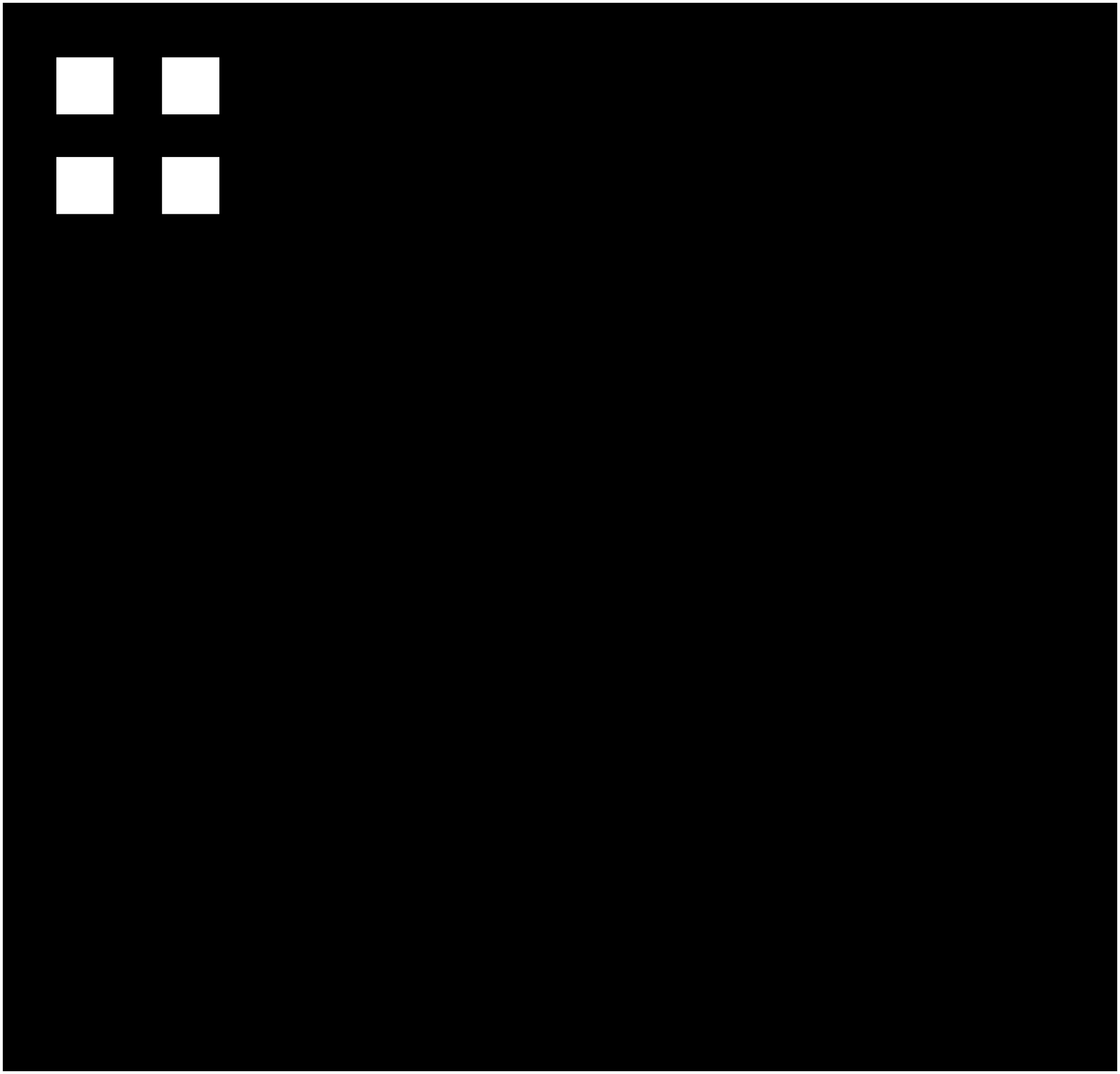}
         \caption{2x2}
         \label{fig:activation:1}
     \end{subfigure}
     \begin{subfigure}[b]{0.15\columnwidth}
         \centering
         \includegraphics[width=\textwidth]{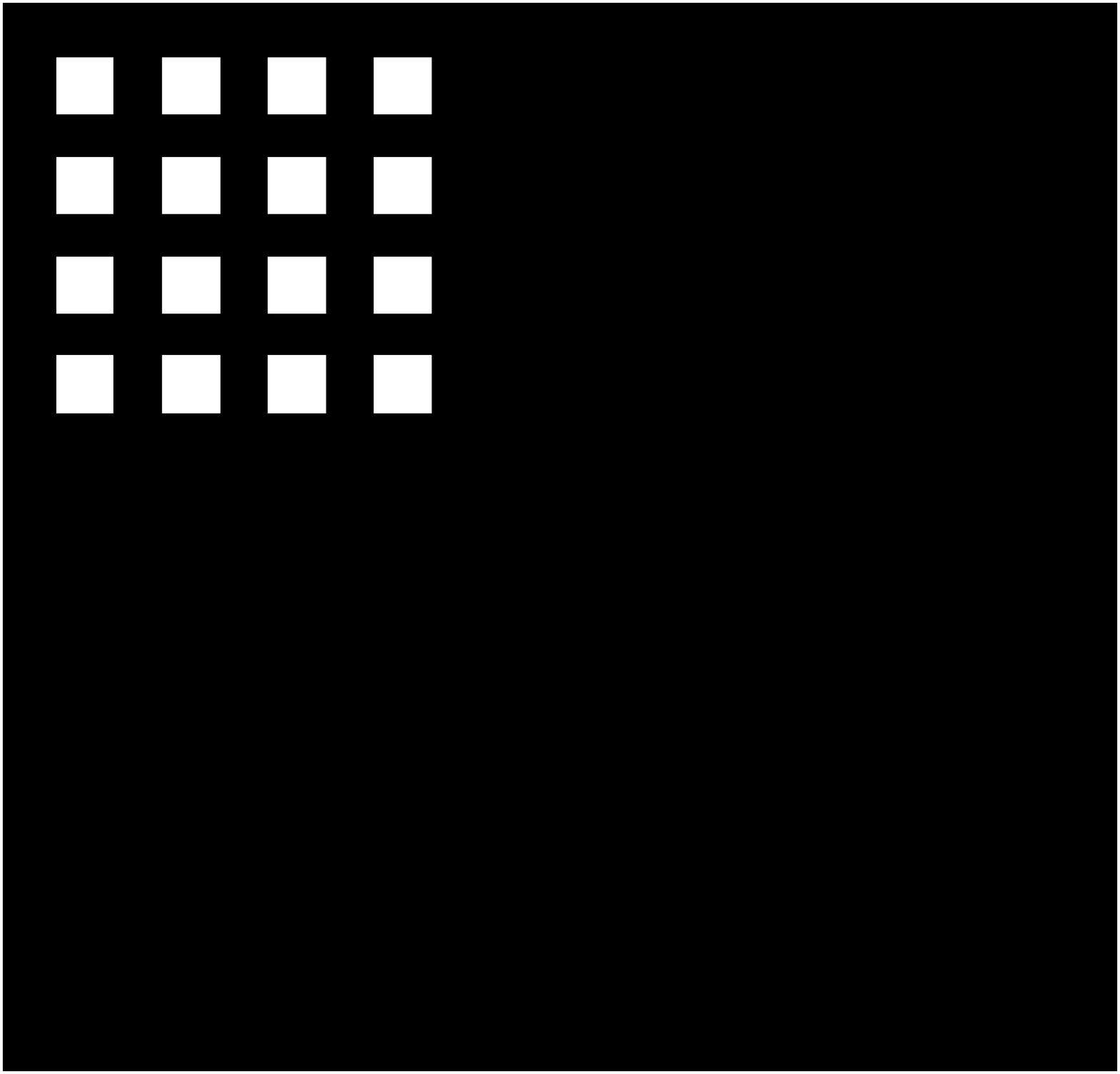}
         \caption{4x4}
         \label{fig:activation:2}
     \end{subfigure}
     \begin{subfigure}[b]{0.15\columnwidth}
         \centering
         \includegraphics[width=\textwidth]{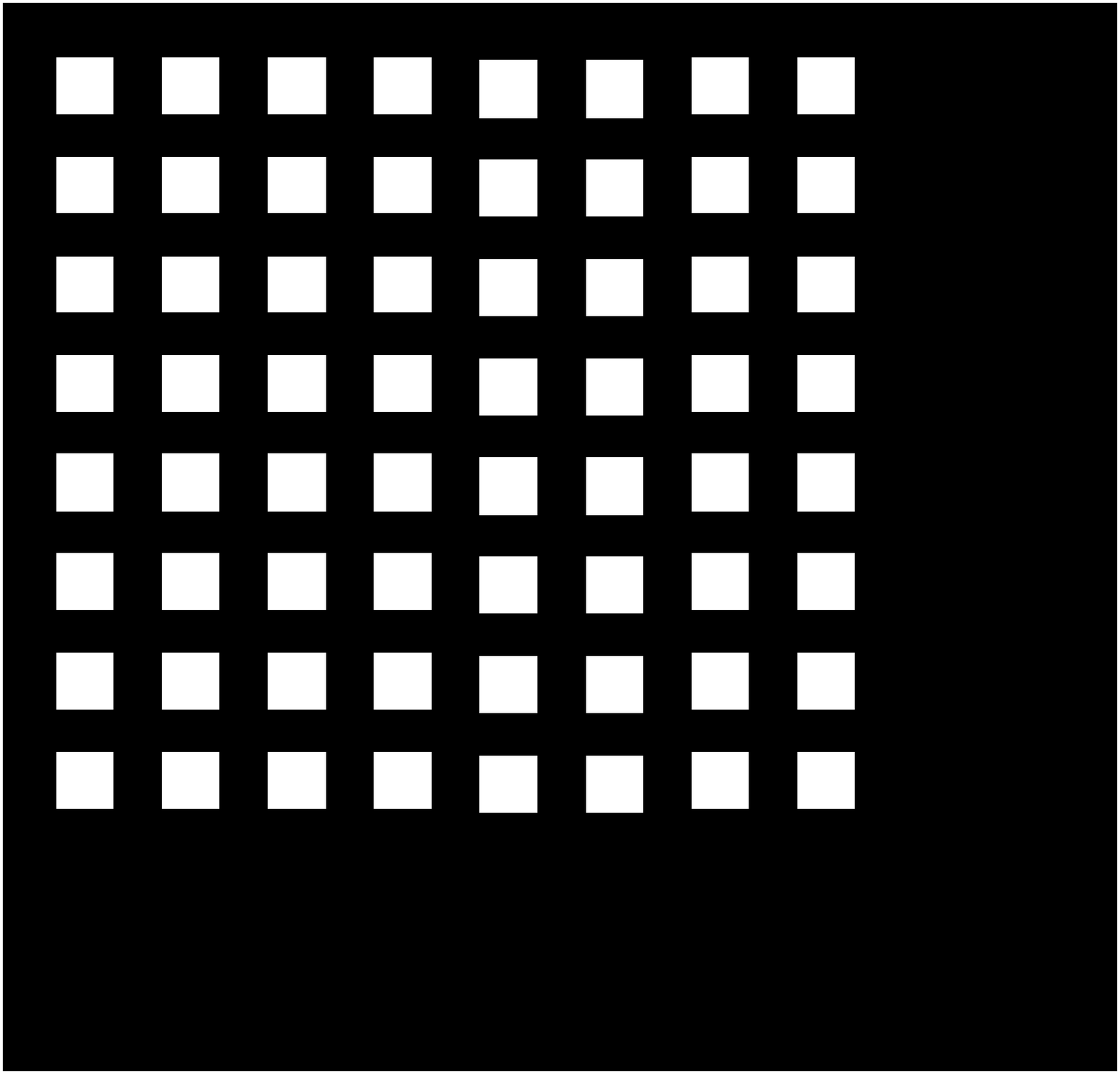}
         \caption{8x8}
         \label{fig:activation:3}
     \end{subfigure}
     \begin{subfigure}[b]{0.15\columnwidth}
         \centering
         \includegraphics[width=\textwidth]{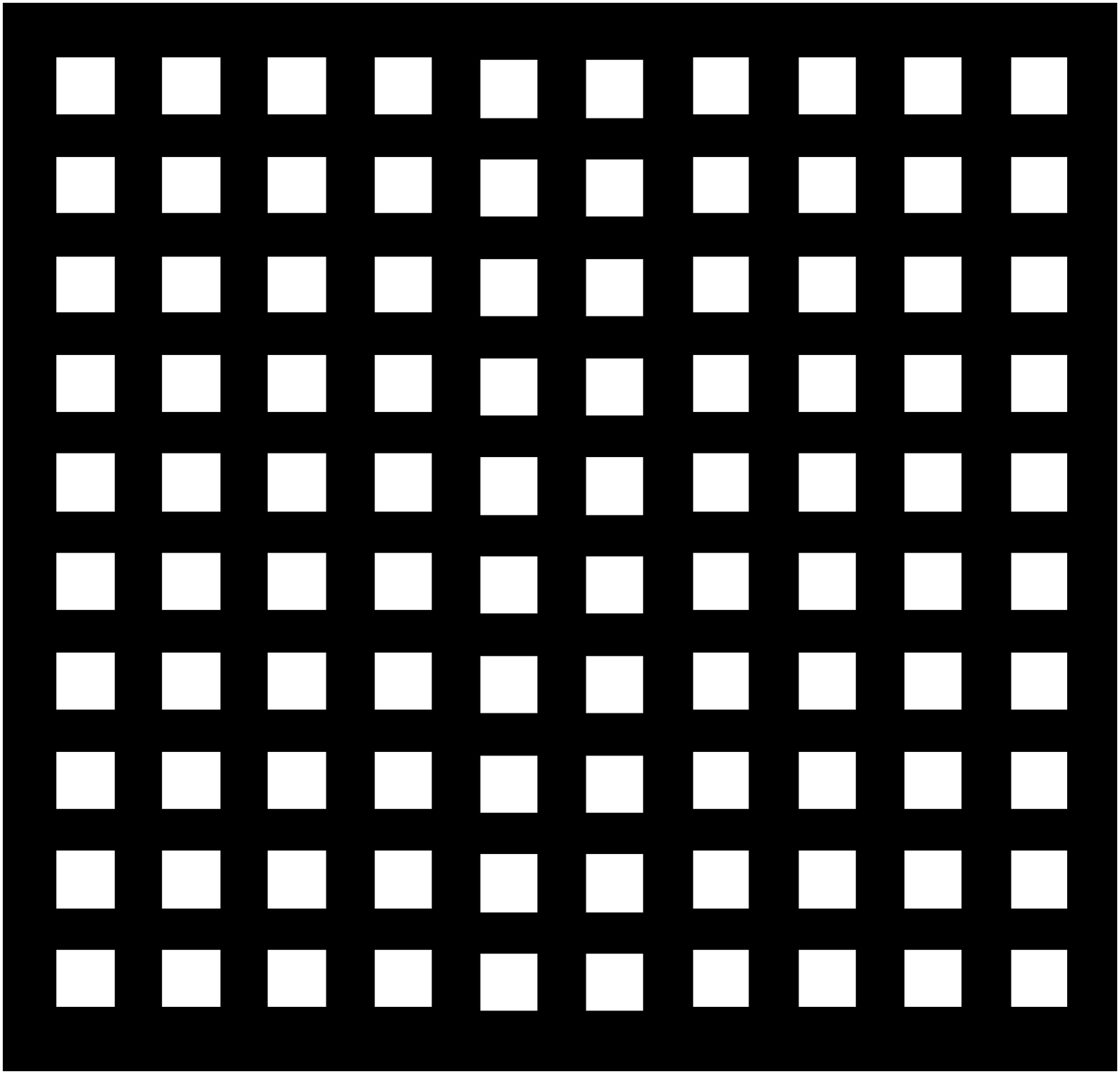}
         \caption{10x10}
         \label{fig:activation:4}
     \end{subfigure}     
     \begin{subfigure}[b]{0.15\columnwidth}
         \centering
         \includegraphics[width=\textwidth]{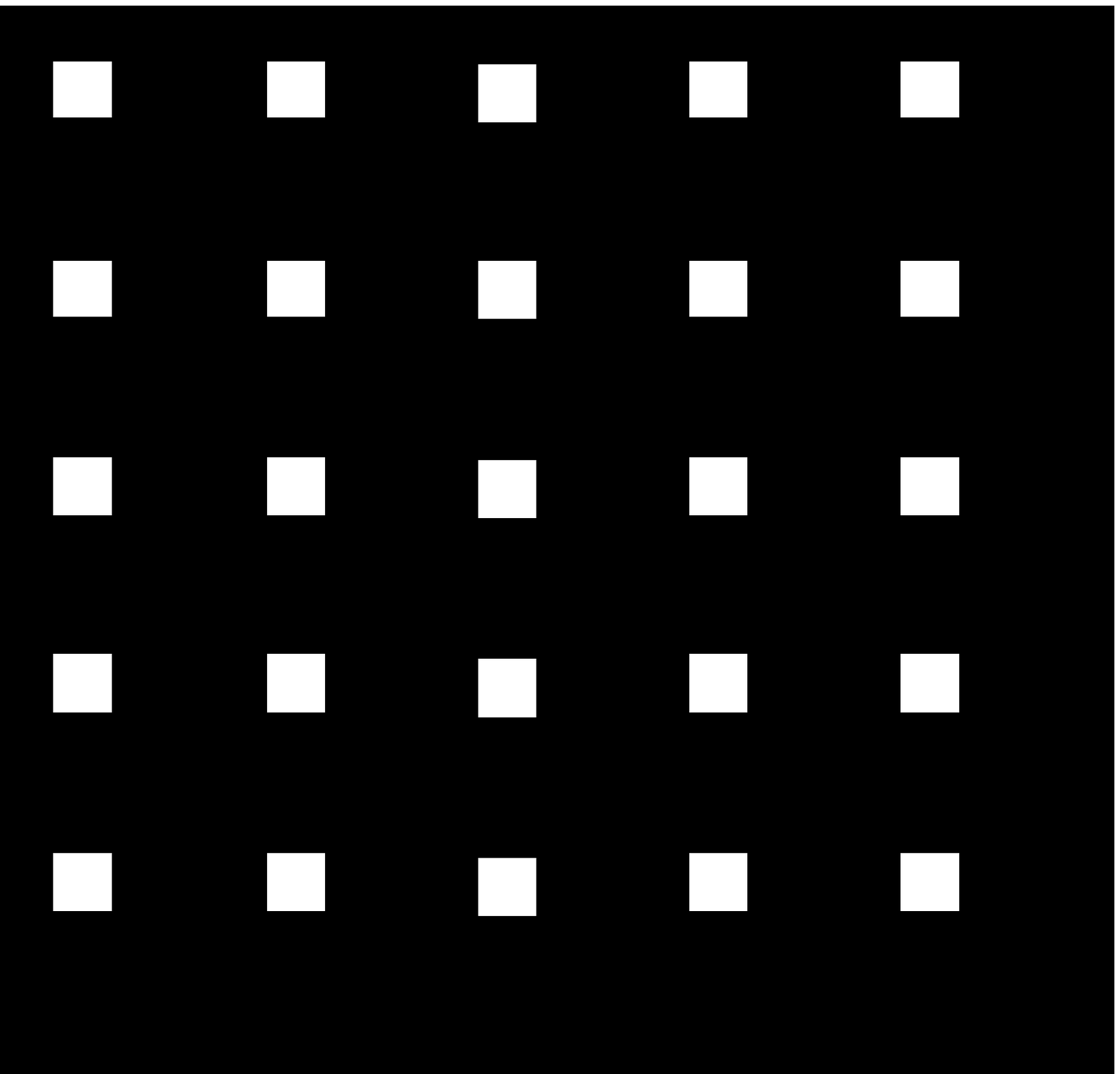}
         \caption{off2}
         \label{fig:activation:5}
     \end{subfigure}  
     \begin{subfigure}[b]{0.15\columnwidth}
         \centering
         \includegraphics[width=\textwidth]{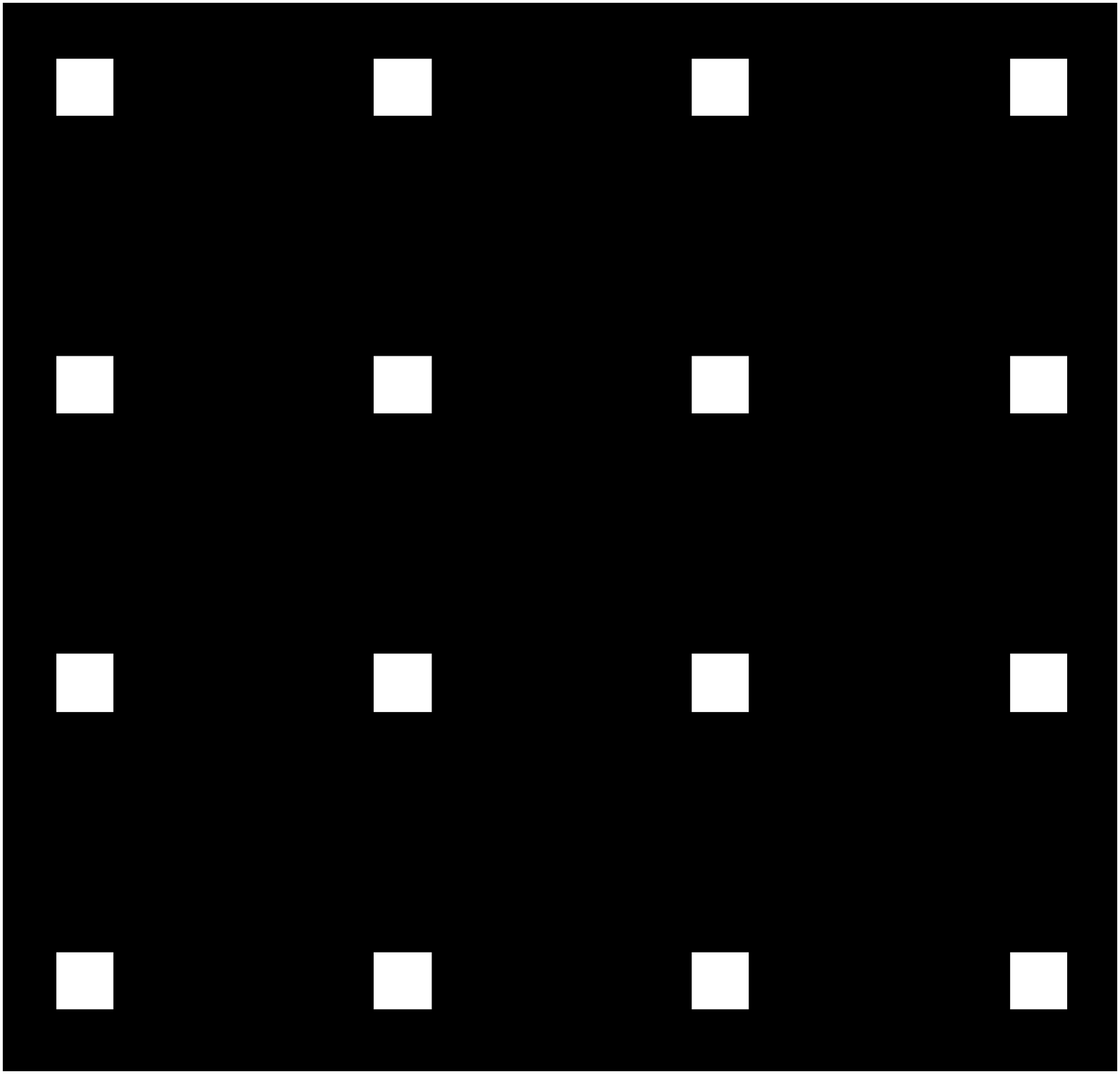}
         \caption{off3}
         \label{fig:activation:6}
     \end{subfigure}     
    \vspace{-4mm}
        \caption{\small Different unit cell activation patterns in a 10x10 RIS board. Colored squares represent deactivated (in black) and activated (in white) cells.}
        \label{fig:activation}
\end{figure}

To assess scalability, we analyze the beamforming gain of our RIS for a variable number of unit cells. To this goal, we take advantage of the  absorption state available in our design  for each cell, and produce virtual RISs with different sizes by setting the activation patterns shown in Figs.~\ref{fig:activation:1} to \ref{fig:activation:4}. For every virtual RIS, we re-optimized its codebook $\mathcal{V}_{N}$ to account for its effective size and inter-cell spacing. 

\begin{figure}[t!]
  \centering
    \includegraphics[width=\columnwidth]{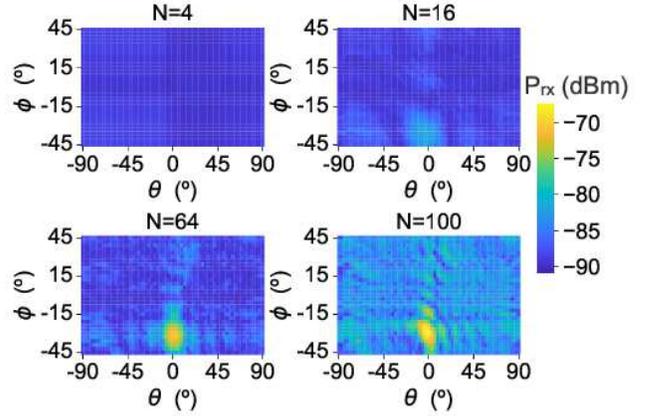}
    \vspace{-8mm}
    \caption{\small Received power over the respective optimal codebook $\boldsymbol{\mathcal{V}_N}$ for virtual RISs with different sizes (activation patterns of Figs.~\ref{fig:activation:1}-d, respectively). $\boldsymbol{\theta_r=\theta_t=0^\circ}$.}
  \label{fig:vRISs}
\end{figure}

To ease the analysis, we now fix $\theta_r=\theta_t=0^\circ$ and measure the power received for every configuration $\v\in\mathcal{V}_N$. The results are shown in Fig.~\ref{fig:vRISs}, which represent the measured power with a color range for every combination of $\theta$ (x-axis) and $\phi$ (y-axis) from $\mathcal{V}_N$, and for $N=\{4, 8, 64, 100\}$. 
From these plots, we can observe how the main beam becomes sharper and carries more power as we increase $N$. With $N=4$ (top left plot), no configuration produces a distinguishable beam, which renders a 2x2 RIS ineffective. For the rest, we note a growing amount of power in the intended direction, respectively, equal to $-81.8$~dBm ($N=16$), $-71.5$~dBm ($N=64$), and $-66.5$~dBm (all cells are activated). This behavior is expected: Fig.~\ref{fig:scalability} depicts in red the power observed at RX with the optimal configuration $\v$ as a function of $N$, and compares that with the mathematical model in \cite{9206044} (see eq.~(5) therein) represented in blue. Both results are remarkably close to each other, which validates the ability of our approach to effectively create virtual surfaces with different shapes.

\subsection{Other activation patterns}

\begin{figure}[b!]
     \centering
     \begin{subfigure}[b]{0.495\columnwidth}
         \centering
         \includegraphics[width=1.05\columnwidth]{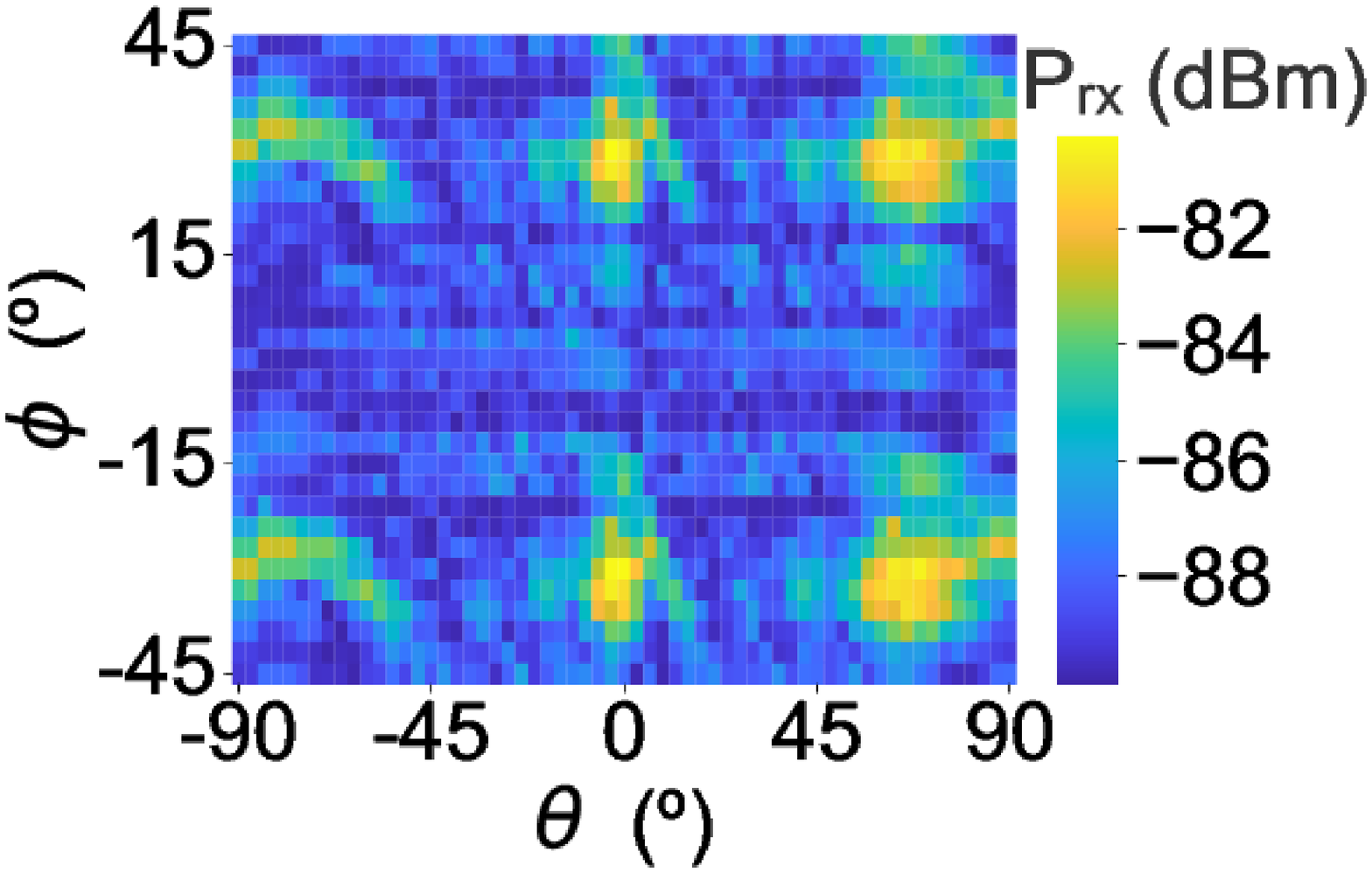}
         \vspace{-7mm}
         \caption{$\boldsymbol{d_{max}=\lambda}$.}
         \label{fig:off2}
     \end{subfigure}  
     \begin{subfigure}[b]{0.495\columnwidth}
         \centering
         \includegraphics[width=1.05\columnwidth]{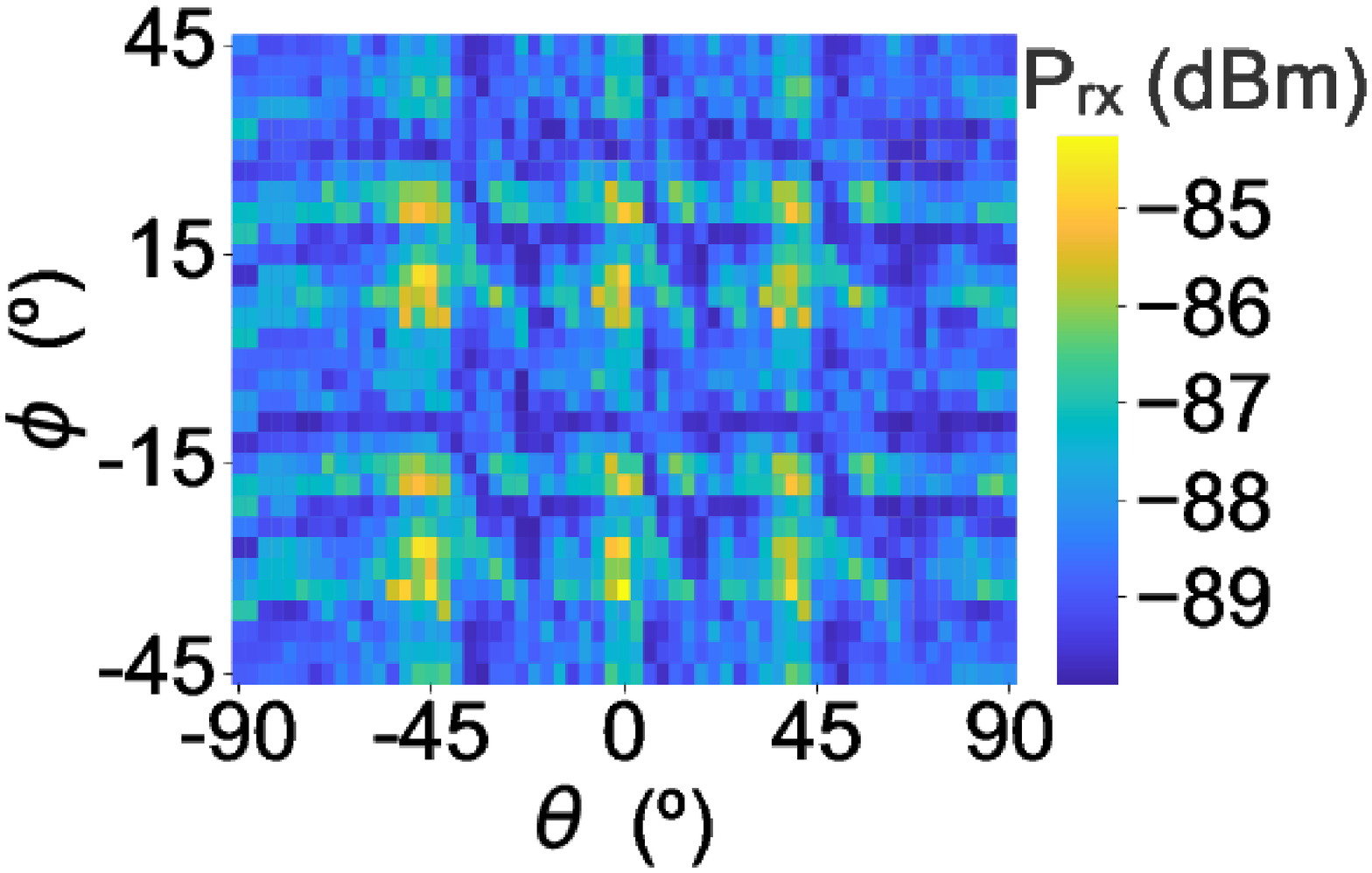}
         \vspace{-7mm}
         \caption{$\boldsymbol{d_{max}=1.5\lambda}$.}
         \label{fig:off3}
     \end{subfigure}     
    \vspace{-6mm}
        \caption{\small Received power over the respective optimal codebook $\boldsymbol{\mathcal{V}_{d_{max}}}$ for virtual RISs generating grating lobes (activation patterns of Figs.~\ref{fig:activation:5}-f, respectively). Red dots indicate the expected location of the lobes. $\boldsymbol{\theta_r=\theta_t=0^\circ}$.}
        \label{fig:grating}
\end{figure}

To conclude our characterization, we study the performance of our RIS prototype when the inter-cell distance differs from $d_{max}=\lambda/2$ (shown in Fig.~\ref{fig:vRISs} for $N=100$).   Like before, we calculated new optimized codebooks $\mathcal{V}_{d_{max}}$ for $d_{max}=\{\lambda, 1.5\lambda\}$, and plot in Fig.~\ref{fig:grating} the power received at RX for each configuration $\v\in\mathcal{V}_{d_{max}}$. We do this for both activation patterns depicted in Fig.~\ref{fig:activation:5} (``off2'') and \ref{fig:activation:6} (``off3''), for $d_{max}=\lambda$ and $d_{max}=1.5\lambda$, respectively. 

By changing $d_{max}$, we also change the density of active cells per board, $N=25$ for $d_{max}=\lambda$ (Fig.~\ref{fig:off2}) and $N=16$ for  $d_{max}=1.5\lambda$ (Fig.~\ref{fig:off3}). As shown earlier, this has a cost in terms of beamforming gains that is evidenced also in Fig.~\ref{fig:grating}: the maximum power is $-80.4$~dBm and $-84.2$~dBm for the two cases, respectively. 
Both plots reveal the presence of grating lobes, which are symmetrical beams that are denser for larger  $d_{max}$ values. 
These effects are well understood in the literature of antenna design and their distance can be estimated using our model in \S\ref{sec:model}. The figure depicts with red circles the expected location of these lobes, which match our measurements remarkably well. This further validates our design to effectively modify the shape of the RIS to the requirements of any given use case. 


\section{Reproducibility}\label{sec:cost}



\begin{figure}[t!]
\minipage{0.48\columnwidth}
    \includegraphics[width=\columnwidth]{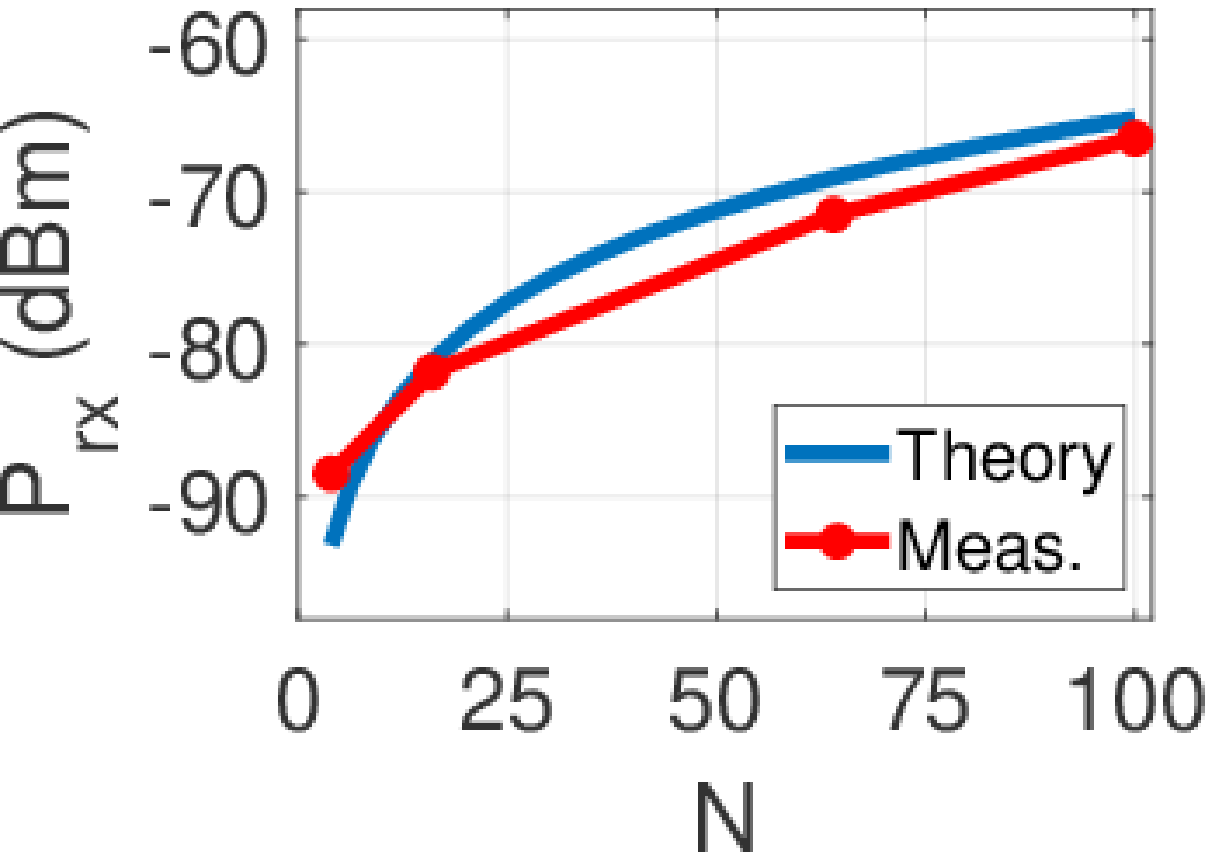}
      \vspace{-8mm}
\caption{\small Beamforming gains as a function of the total number of antennas $N$.}
    \label{fig:scalability}
\endminipage
\hfill
\minipage{0.48\columnwidth}
  \includegraphics[width=\columnwidth]{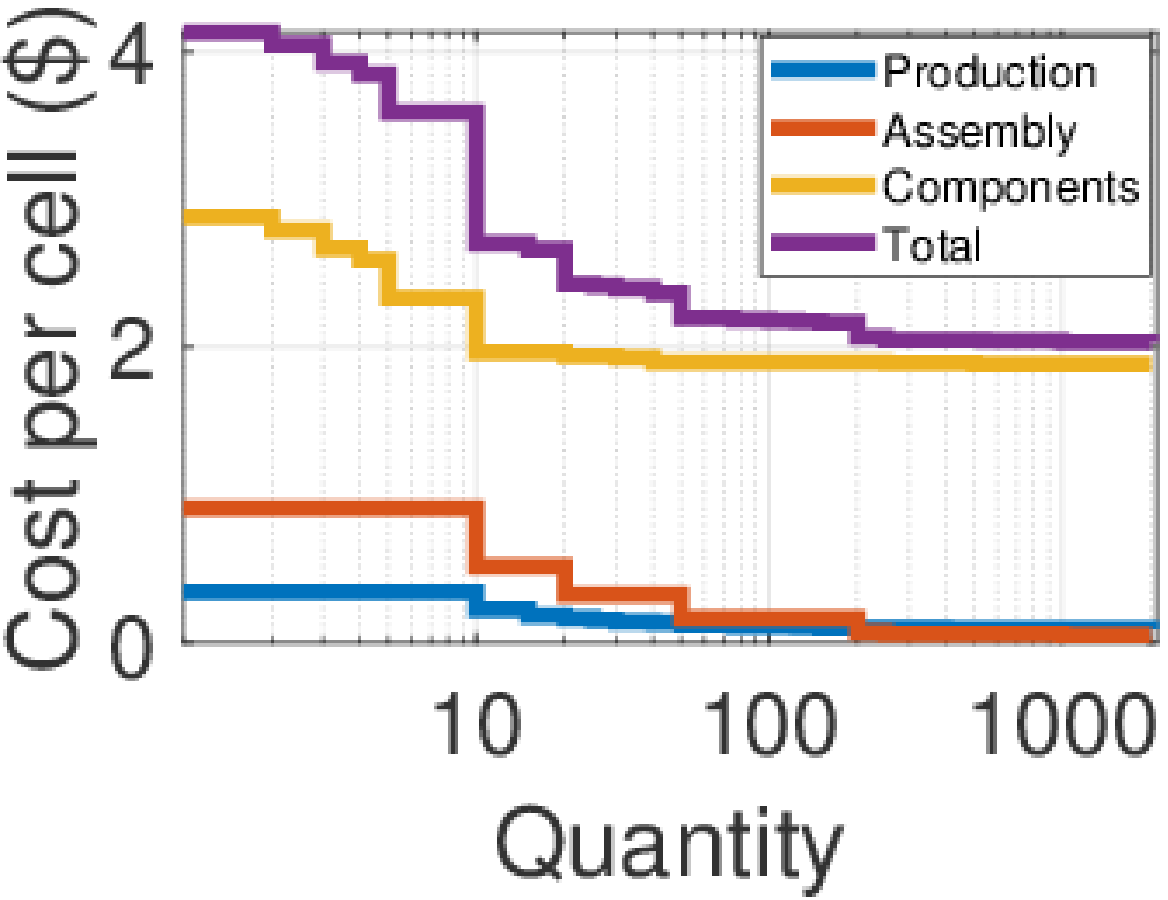}
    \vspace{-6mm}
  \caption{\small Prototype costs over manufacturing scale.}
  \label{fig:cost}
\endminipage
\end{figure}

To conclude our paper, we provide some final remarks that shall help researchers in the RIS domain build on our results (dataset) and/or reproduce our RIS prototype. 

For starters, we publicly release the dataset\footnote{We will publish the dataset upon the paper acceptance.} we have generated during our empirical characterization in \S\ref{sec:results}, which aggregates a total number of $6.8\cdot 10^6$ power samples. This data can help other researchers study RIS-related problems without the need of building a prototype. 

Next, we report the costs associated for building our prototype. For this purpose, we have used prices publicly available. 
There are three sources of cost to manufacture each board: ($i$) PCB production (including the patch antenna and the microstrips for the delay lines and the buses); ($ii$) additional electronic components (including RF switches, MCU, resistors, flip-flops, AND gates, etc.); and ($iii$) assembly all components onto the PCB. 

According to {\ttfamily PCBWay}~\cite{PCBWay}, producing 10 PCB boards, as specified in our design, costs \$0.22 per unit cell; but it drops to \$0.11 and \$0.09 when the production scales up by 20x and 100x, respectively. 
Concerning additional electronic components, each 10x10 board bears 300 flip-flops, 100 AND gates, 100 RF switches, and 1 MCU. According to {\ttfamily Digikey.de}, the cost boils down to \$1.88 per unit cell when purchasing a batch of 1000 of such boards.
The assembly process, according to PCBWay, scales down from \$0.51 to \$0.05 per cell when we scale up the number of boards from 10 to 1000 10x10-RISs. 
Fig.~\ref{fig:cost} depicts how these costs (normalized per unit cell) evolve with the manufacturing scale. 

\section{Conclusion}\label{sec:conclusions}
Reconfigurable Intelligent Surfaces (RISs) can become a major densification solution for future mobile systems. We designed, prototyped and characterized an RF switch-based RIS that met our defined requirements in terms of RF steering, reconfigurability, energy efficiency, and cost. 
Throughout the paper, we reported on our learnings along with the related experiments. These learnings ranged from our experience bridging theory and empirical findings (e.g., unexpected sensitivity to certain model parameters) to practical considerations (e.g., costs and hardware constrains).
As a result, our contributions can be used by the research community to: ($i$) build realistic RF switch-based RIS models, ($ii$) leverage on the dataset provided to further study related challenges, ($iii$) reproduce our RIS prototype for research purposes, and ($iv$) estimate the deployment costs at scale.





\begin{thebibliography}{24}


\ifx \showCODEN    \undefined \def \showCODEN     #1{\unskip}     \fi
\ifx \showDOI      \undefined \def \showDOI       #1{#1}\fi
\ifx \showISBNx    \undefined \def \showISBNx     #1{\unskip}     \fi
\ifx \showISBNxiii \undefined \def \showISBNxiii  #1{\unskip}     \fi
\ifx \showISSN     \undefined \def \showISSN      #1{\unskip}     \fi
\ifx \showLCCN     \undefined \def \showLCCN      #1{\unskip}     \fi
\ifx \shownote     \undefined \def \shownote      #1{#1}          \fi
\ifx \showarticletitle \undefined \def \showarticletitle #1{#1}   \fi
\ifx \showURL      \undefined \def \showURL       {\relax}        \fi
\providecommand\bibfield[2]{#2}
\providecommand\bibinfo[2]{#2}
\providecommand\natexlab[1]{#1}
\providecommand\showeprint[2][]{arXiv:#2}

\bibitem[\protect\citeauthoryear{Albanese et~al\mbox{.}}{Albanese
  et~al\mbox{.}}{2022}]%
        {albanese22}
\bibfield{author}{\bibinfo{person}{Antonio Albanese} {et~al\mbox{.}}}
  \bibinfo{year}{2022}\natexlab{}.
\newblock \showarticletitle{{MARISA: A Self-configuring Metasurfaces Absorption
  and Reflection Solution Towards 6G}}. In \bibinfo{booktitle}{\emph{IEEE
  INFOCOM 2022 - IEEE Conference on Computer Communications}}.
\newblock


\bibitem[\protect\citeauthoryear{{Analog Devices}}{{Analog Devices}}{2009}]%
        {strip}
\bibfield{author}{\bibinfo{person}{{Analog Devices}}.}
  \bibinfo{year}{2009}\natexlab{}.
\newblock \bibinfo{title}{{Microstrip and Stripline Design: MT-094 tutorial}}.
\newblock
  \bibinfo{howpublished}{\url{https://www.analog.com/media/en/training-seminars/tutorials/MT-094.pdf}}.
    (\bibinfo{year}{2009}).
\newblock
\newblock
\shownote{Accessed: 2022-03-01.}


\bibitem[\protect\citeauthoryear{Arun and Balakrishnan}{Arun and
  Balakrishnan}{2020}]%
        {arun2020rfocus}
\bibfield{author}{\bibinfo{person}{Venkat Arun} {and} \bibinfo{person}{Hari
  Balakrishnan}.} \bibinfo{year}{2020}\natexlab{}.
\newblock \showarticletitle{RFocus: Beamforming using thousands of passive
  antennas}. In \bibinfo{booktitle}{\emph{USENIX NSDI}}.
\newblock


\bibitem[\protect\citeauthoryear{Balanis}{Balanis}{2016}]%
        {balanis}
\bibfield{author}{\bibinfo{person}{Constantine~A Balanis}.}
  \bibinfo{year}{2016}\natexlab{}.
\newblock \bibinfo{booktitle}{\emph{Antenna theory: analysis and design}}.
\newblock \bibinfo{publisher}{John wiley \& sons}.
\newblock


\bibitem[\protect\citeauthoryear{Basar et~al\mbox{.}}{Basar
  et~al\mbox{.}}{2019}]%
        {basar2019wireless}
\bibfield{author}{\bibinfo{person}{Ertugrul Basar} {et~al\mbox{.}}}
  \bibinfo{year}{2019}\natexlab{}.
\newblock \showarticletitle{Wireless communications through reconfigurable
  intelligent surfaces}.
\newblock \bibinfo{journal}{\emph{IEEE access}}  \bibinfo{volume}{7}
  (\bibinfo{year}{2019}), \bibinfo{pages}{116753--116773}.
\newblock


\bibitem[\protect\citeauthoryear{Dai et~al\mbox{.}}{Dai et~al\mbox{.}}{2020}]%
        {dai2020reconfigurable}
\bibfield{author}{\bibinfo{person}{Linglong Dai} {et~al\mbox{.}}}
  \bibinfo{year}{2020}\natexlab{}.
\newblock \showarticletitle{Reconfigurable intelligent surface-based wireless
  communications: Antenna design, prototyping, and experimental results}.
\newblock \bibinfo{journal}{\emph{IEEE Access}}  \bibinfo{volume}{8}
  (\bibinfo{year}{2020}), \bibinfo{pages}{45913--45923}.
\newblock


\bibitem[\protect\citeauthoryear{{Dassault Syst\`{e}mes}}{{Dassault
  Syst\`{e}mes}}{2022}]%
        {cststudio}
\bibfield{author}{\bibinfo{person}{{Dassault Syst\`{e}mes}}.}
  \bibinfo{year}{2022}\natexlab{}.
\newblock \bibinfo{title}{{CST Studio Suite: Electromagnetic field simulation
  software}}.
\newblock
  \bibinfo{howpublished}{\url{https://www.3ds.com/products-services/simulia/products/cst-studio-suite/}}.
    (\bibinfo{year}{2022}).
\newblock
\newblock
\shownote{Accessed: 2022-03-01.}


\bibitem[\protect\citeauthoryear{Delos, Broughton, and Kraft}{Delos
  et~al\mbox{.}}{2020}]%
        {AD}
\bibfield{author}{\bibinfo{person}{Peter Delos}, \bibinfo{person}{Bob
  Broughton}, {and} \bibinfo{person}{Jon Kraft}.}
  \bibinfo{year}{2020}\natexlab{}.
\newblock \showarticletitle{Phased Array Antenna Patterns—Part 2: Grating
  Lobes and Beam Squint}.
\newblock \bibinfo{journal}{\emph{26 Phased Array Antenna Patterns—Series}}
  (\bibinfo{year}{2020}), \bibinfo{pages}{39}.
\newblock


\bibitem[\protect\citeauthoryear{Dunna et~al\mbox{.}}{Dunna
  et~al\mbox{.}}{2020}]%
        {scattermimo}
\bibfield{author}{\bibinfo{person}{Manideep Dunna} {et~al\mbox{.}}}
  \bibinfo{year}{2020}\natexlab{}.
\newblock \showarticletitle{ScatterMIMO: Enabling virtual MIMO with smart
  surfaces}. In \bibinfo{booktitle}{\emph{ACM MobiCom}}.
\newblock


\bibitem[\protect\citeauthoryear{Fara et~al\mbox{.}}{Fara
  et~al\mbox{.}}{2021}]%
        {RomainDiRenzo2021}
\bibfield{author}{\bibinfo{person}{Romain Fara} {et~al\mbox{.}}}
  \bibinfo{year}{2021}\natexlab{}.
\newblock \showarticletitle{A Prototype of Reconfigurable Intelligent Surface
  with Continuous Control of the Reflection Phase}.
\newblock \bibinfo{journal}{\emph{arXiv preprint arXiv:2105.11862}}
  (\bibinfo{year}{2021}).
\newblock


\bibitem[\protect\citeauthoryear{Hu et~al\mbox{.}}{Hu et~al\mbox{.}}{2020}]%
        {VincentPoor2020}
\bibfield{author}{\bibinfo{person}{Jingzhi Hu} {et~al\mbox{.}}}
  \bibinfo{year}{2020}\natexlab{}.
\newblock \showarticletitle{Reconfigurable Intelligent Surface Based RF
  Sensing: Design, Optimization, and Implementation}.
\newblock \bibinfo{journal}{\emph{IEEE Journal on Selected Areas in
  Communications}} \bibinfo{volume}{38}, \bibinfo{number}{11}
  (\bibinfo{year}{2020}), \bibinfo{pages}{2700--2716}.
\newblock


\bibitem[\protect\citeauthoryear{Mir et~al\mbox{.}}{Mir et~al\mbox{.}}{2021}]%
        {mir2021passivelifi}
\bibfield{author}{\bibinfo{person}{Muhammad~Sarmad Mir} {et~al\mbox{.}}}
  \bibinfo{year}{2021}\natexlab{}.
\newblock \showarticletitle{PassiveLiFi: rethinking LiFi for low-power and long
  range RF backscatter}. In \bibinfo{booktitle}{\emph{ACM MobiCom}}.
\newblock


\bibitem[\protect\citeauthoryear{Mursia et~al\mbox{.}}{Mursia
  et~al\mbox{.}}{2021}]%
        {RISMA}
\bibfield{author}{\bibinfo{person}{Placido Mursia} {et~al\mbox{.}}}
  \bibinfo{year}{2021}\natexlab{}.
\newblock \showarticletitle{{RISMA: Reconfigurable Intelligent Surfaces
  Enabling Beamforming for IoT Massive Access}}.
\newblock \bibinfo{journal}{\emph{IEEE Journal on Selected Areas in
  Communications}} \bibinfo{volume}{39}, \bibinfo{number}{4}
  (\bibinfo{year}{2021}), \bibinfo{pages}{1072--1085}.
\newblock


\bibitem[\protect\citeauthoryear{Pan et~al\mbox{.}}{Pan et~al\mbox{.}}{2021}]%
        {9475160}
\bibfield{author}{\bibinfo{person}{Cunhua Pan} {et~al\mbox{.}}}
  \bibinfo{year}{2021}\natexlab{}.
\newblock \showarticletitle{Reconfigurable Intelligent Surfaces for 6G Systems:
  Principles, Applications, and Research Directions}.
\newblock \bibinfo{journal}{\emph{IEEE Communications Magazine}}
  \bibinfo{volume}{59}, \bibinfo{number}{6} (\bibinfo{year}{2021}),
  \bibinfo{pages}{14--20}.
\newblock


\bibitem[\protect\citeauthoryear{Pandey}{Pandey}{2019}]%
        {pandey2019practical}
\bibfield{author}{\bibinfo{person}{Anil Pandey}.}
  \bibinfo{year}{2019}\natexlab{}.
\newblock \bibinfo{booktitle}{\emph{Practical microstrip and printed antenna
  design}}.
\newblock \bibinfo{publisher}{Artech House}.
\newblock


\bibitem[\protect\citeauthoryear{{PCBWay}}{{PCBWay}}{[n. d.]}]%
        {PCBWay}
\bibfield{author}{\bibinfo{person}{{PCBWay}}.} \bibinfo{year}{[n.
  d.]}\natexlab{}.
\newblock \bibinfo{title}{{PCB prototype and fabrication}}.
\newblock \bibinfo{howpublished}{\url{https://www.pcbway.com}}.
  (\bibinfo{year}{[n. d.]}).
\newblock


\bibitem[\protect\citeauthoryear{{Skyworks Solutions, Inc}}{{Skyworks
  Solutions, Inc}}{2019}]%
        {SKY}
\bibfield{author}{\bibinfo{person}{{Skyworks Solutions, Inc}}.}
  \bibinfo{year}{2019}\natexlab{}.
\newblock \bibinfo{title}{{SKY13418-485LF: 0.1 to 6.0 GHz SP8T Antenna
  Switch}}.
\newblock
  \bibinfo{howpublished}{\url{https://www.skyworksinc.com/-/media/SkyWorks/Documents/Products/701-800/SKY13418_485LF_201712F.pdf}}.
    (\bibinfo{year}{2019}).
\newblock
\newblock
\shownote{Accessed: 2022-03-01.}


\bibitem[\protect\citeauthoryear{{STMicroelectronics}}{{STMicroelectronics}}{2021}]%
        {mcu}
\bibfield{author}{\bibinfo{person}{{STMicroelectronics}}.}
  \bibinfo{year}{2021}\natexlab{}.
\newblock \bibinfo{title}{{STM32L071V8: Ultra-low-power Arm Cortex-M0+ MCU with
  64-Kbytes of Flash memory, 32 MHz CPU }}.
\newblock
  \bibinfo{howpublished}{\url{https://www.st.com/en/microcontrollers-microprocessors/stm32l071v8.html}}.
    (\bibinfo{year}{2021}).
\newblock
\newblock
\shownote{Accessed: 2022-03-01.}


\bibitem[\protect\citeauthoryear{Tan, Sun, Koutsonikolas, and Jornet}{Tan
  et~al\mbox{.}}{2018}]%
        {tan2018enabling}
\bibfield{author}{\bibinfo{person}{Xin Tan}, \bibinfo{person}{Zhi Sun},
  \bibinfo{person}{Dimitrios Koutsonikolas}, {and} \bibinfo{person}{Josep~M
  Jornet}.} \bibinfo{year}{2018}\natexlab{}.
\newblock \showarticletitle{Enabling indoor mobile millimeter-wave networks
  based on smart reflect-arrays}. In \bibinfo{booktitle}{\emph{IEEE INFOCOM
  2018-IEEE Conference on Computer Communications}}. IEEE,
  \bibinfo{pages}{270--278}.
\newblock


\bibitem[\protect\citeauthoryear{Tang et~al\mbox{.}}{Tang
  et~al\mbox{.}}{2021}]%
        {9206044}
\bibfield{author}{\bibinfo{person}{Wankai Tang} {et~al\mbox{.}}}
  \bibinfo{year}{2021}\natexlab{}.
\newblock \showarticletitle{Wireless Communications With Reconfigurable
  Intelligent Surface: Path Loss Modeling and Experimental Measurement}.
\newblock \bibinfo{journal}{\emph{IEEE Transactions on Wireless
  Communications}} \bibinfo{volume}{20}, \bibinfo{number}{1}
  (\bibinfo{year}{2021}), \bibinfo{pages}{421--439}.
\newblock
\urldef\tempurl%
\url{https://doi.org/10.1109/TWC.2020.3024887}
\showDOI{\tempurl}


\bibitem[\protect\citeauthoryear{Trichopoulos et~al\mbox{.}}{Trichopoulos
  et~al\mbox{.}}{2021}]%
        {trichopoulos2021design}
\bibfield{author}{\bibinfo{person}{Georgios Trichopoulos} {et~al\mbox{.}}}
  \bibinfo{year}{2021}\natexlab{}.
\newblock \showarticletitle{Design and Evaluation of Reconfigurable Intelligent
  Surfaces in Real-World Environment}.
\newblock \bibinfo{journal}{\emph{arXiv preprint arXiv:2109.07763}}
  (\bibinfo{year}{2021}).
\newblock


\bibitem[\protect\citeauthoryear{Wu et~al\mbox{.}}{Wu et~al\mbox{.}}{2018}]%
        {ZhangActivePassive}
\bibfield{author}{\bibinfo{person}{Qingqing Wu} {et~al\mbox{.}}}
  \bibinfo{year}{2018}\natexlab{}.
\newblock \showarticletitle{{Intelligent Reflecting Surface Enhanced Wireless
  Network: Joint Active and Passive Beamforming Design}}.
\newblock \bibinfo{journal}{\emph{IEEE Transactions on Wireless
  Communications}} \bibinfo{volume}{18}, \bibinfo{number}{11}
  (\bibinfo{year}{2018}), \bibinfo{pages}{5394--5409}.
\newblock


\bibitem[\protect\citeauthoryear{Yang et~al\mbox{.}}{Yang
  et~al\mbox{.}}{2022}]%
        {9497709}
\bibfield{author}{\bibinfo{person}{Zhaohui Yang} {et~al\mbox{.}}}
  \bibinfo{year}{2022}\natexlab{}.
\newblock \showarticletitle{Energy-Efficient Wireless Communications With
  Distributed Reconfigurable Intelligent Surfaces}.
\newblock \bibinfo{journal}{\emph{IEEE Transactions on Wireless
  Communications}} \bibinfo{volume}{21}, \bibinfo{number}{1}
  (\bibinfo{year}{2022}), \bibinfo{pages}{665--679}.
\newblock


\bibitem[\protect\citeauthoryear{Yezhen et~al\mbox{.}}{Yezhen
  et~al\mbox{.}}{2020}]%
        {yezhen2020novel}
\bibfield{author}{\bibinfo{person}{Li Yezhen} {et~al\mbox{.}}}
  \bibinfo{year}{2020}\natexlab{}.
\newblock \showarticletitle{A Novel 28 GHz Phased Array Antenna for 5G Mobile
  Communications}.
\newblock \bibinfo{journal}{\emph{ZTE Communications}} \bibinfo{volume}{18},
  \bibinfo{number}{3} (\bibinfo{year}{2020}), \bibinfo{pages}{20--25}.
\newblock


\end{thebibliography}

\end{document}